\documentclass[a4paper,fleqn]{cas-dc}
\usepackage[numbers,sort&compress]{natbib}

\usepackage[utf8]{inputenc}
\usepackage{amsmath,amssymb}
\usepackage{tabularx}
\usepackage{enumitem}
\usepackage{longtable}
\usepackage{xcolor}
\usepackage{array}
\usepackage{booktabs}
\usepackage{makecell}
\usepackage{colortbl}
\usepackage{threeparttable}
\usepackage{placeins}
\usepackage{tikz}
\usepackage{multirow}
\usepackage{lineno}
\usepackage{hyperref}
\usepackage{xurl}
\usepackage{xfp}
\usepackage{stfloats}
\usepackage{caption} 
\usepackage{doi}  
\usepackage{hyphenat}
\usepackage{fontawesome5}
\usetikzlibrary{positioning, arrows.meta, fit, backgrounds, calc}

\makeatletter
\@ifundefined{affiliation}{\newcommand{\affiliation}[2][]{}}{}
\makeatother

\begin{document}

\let\WriteBookmarks\relax
\def\floatpagepagefraction{1}
\def\textpagefraction{.001}
\renewcommand{\printorcid}{}

\definecolor{socmint}{HTML}{0EA5E9}
\definecolor{humint}{HTML}{7C3AED}
\definecolor{phasebg}{HTML}{1E293B}
\definecolor{hdrbg}{RGB}{58,58,58}
\definecolor{rowalt}{RGB}{248,248,250}

\shorttitle{Social-HUMINT: An LLM-Augmented Unified Framework for Measuring Social Media Exposure} 

\shortauthors{Casoria and Romano}

\title [mode = title]{Social-HUMINT: An LLM-Augmented Unified Framework for Measuring Social Media Exposure}  

\author[1]{Pellegrino Casoria}%
\cormark[1]                         
\ead{pellegrino.casoria@unina.it} 
\credit{Conceptualization, Data curation, Formal analysis, Investigation, Methodology, Project administration, Resources, Software, Validation, Visualization, Writing -- original draft, Writing -- review and editing}  

\author[1]{Simon Pietro Romano}%
\credit{Project administration, Supervision, Writing -- review and editing}

\cortext[1]{Corresponding author}

\affiliation[1]{organization={University of Naples Federico II},
                department={Department of Electrical Engineering and Information Technology (DIETI),},
                city={Naples},
                country={Italy}}

\begin{abstract}
In the post-API landscape, social-media profiling has become inexpensive and automated: Large Language Models (LLMs) can extract structured information from unstructured content and deduce attributes users never disclosed. A person's true exposure is not what they post but what can be inferred from it. As these practices become routine, exposure to social engineering attacks becomes an operational certainty, demanding a shared framework for assessment and mitigation.

We make this exposure measurable by quantifying how accurately a target can be profiled, and how far that profile enables social engineering attacks. Because no existing model fully spans this continuum, we unify Social Media Intelligence (SOCMINT) and Cyber Human Intelligence (Cyber-HUMINT), two individual-focused disciplines that evolved separately, into a single doctrine-grounded taxonomy of 38 capabilities across the Cyber Threat Intelligence lifecycle, where user profiling bridges passive collection and active engagement. Domain experts validated the taxonomy (Fleiss' $\kappa = 0.92$).

We operationalize its core path (extraction, inference, and susceptibility assessment) as an LLM-augmented multi-agent pipeline, evaluated on real participants' public data. Susceptibility is assessed indirectly: the pipeline predicts how each target would rank personalized engagement options, matching their actual choices significantly above chance. This ranking capacity captures the personalization on which tailored attacks depend. The underlying profiles are accurate: extraction reaches $F_1 = 97.3\%$, and inferred attributes score $4.73/5$, both far above rule-based and NER-based baselines at negligible cost. The framework augments security analysts' tradecraft rather than replacing it and enables defenders to gauge their own exposure before an adversary does.
\end{abstract}


\maketitle

\section{Introduction} 
\label{sec:introduction}

Social media has transformed how people communicate and share information. It now has about 5.66 billion active users (over 68\% of the global population), grows at 4.1\% per year, and accounts for more than 2 hours of daily use across an average of 7 platforms \cite{datareportal2025social,demandsage2025social}. These public footprints expose users to security risks that extend beyond what they post. Beneath visible content lies an inferred layer (organizational access, behavioral traits, communication style) that users never disclose but that can still be reconstructed from sparse public signals \cite{staab_beyond_2024}. This inferred layer is part of a person's true exposure, yet current practice neither observes nor measures it.

The dual-use nature of this environment is evident in current tradecraft. MITRE ATT\&CK documents systematic reconnaissance through publicly available social data~\cite{mitre2024t1593}. APT campaigns such as IN(TER)CEPTION~\cite{mitre_lazarus}, Dream Job~\cite{mitre_dream_job}, EXOTIC LILY~\cite{google_exotic_lily}, and Contagious Interview~\cite{mitre_contagious_interview} begin with counterfeit personas, cloned profiles, and social engineering on LinkedIn and other platforms before any technical intrusion. This trend is accelerating. IBM X-Force reports an 84\% increase in phishing campaigns delivering infostealers identified through social media monitoring \cite{ibm2025xforce}. CrowdStrike reports faster adversary breakout times and malware-free techniques that are often first observable in social media discourse \cite{crowdstrike2025threat}.

The Cambridge Analytica scandal \cite{cadwalladr2018cambridge} marked the start of the post-API era. Platforms restricted their APIs, while GDPR and the Digital Services Act constrained lawful collection \cite{bruns2019apiocalypse,gdpr2016regulation}. As a result, profiles can now be built only by extracting data from unstructured, platform-specific sources, often through scraping that raises Terms-of-Service and privacy concerns \cite{trezza2023scraping}. Platforms also enforce defenses through access policies and data-protection measures. Yet these controls are increasingly fragile. Large Language Models (LLMs) can extract structured information from unstructured sources at scale and infer latent attributes that platforms never explicitly exposed.

This attack surface is hard to measure because most steps that turn observable data into exploitable intelligence happen out of sight, before any technical intrusion. These invisible stages can be reconstructed only by mirroring the adversary through legitimate disciplines built on the same capabilities. SOCMINT (Social Media Intelligence) \cite{omand2012socmint} supports passive, observation-driven collection and analysis of social media data. Cyber-HUMINT relies on active engagement, with operators interacting with targets to elicit what observation cannot reveal. Both are already operational, from government SOCMINT programs that monitor extremist recruitment and foreign influence to Cyber-HUMINT operations that infiltrate ransomware groups and darknet marketplaces \cite{fbi2022oversight,brennan2023surveillance,dimaggio2023lockbit,groupib2025}.

Both disciplines focus on the same individuals and differ mainly in depth of engagement, yet they have evolved as separate workflows. Existing models cover only parts of this path. Kill-chain and ATT\&CK-style taxonomies enumerate adversary phases \cite{lockheed2011killchain,mitre2024t1593}, and OSINT frameworks structure passive collection \cite{pastor-galindo_not_2020}. However, none formalizes the passive-to-active continuum at the level of operational capabilities, where otherwise invisible exposure becomes observable. How far these capabilities enable attacks, and how much of that path LLMs can automate, remain open empirical questions that require direct study.

\begin{figure}
    \centering
    \includegraphics[width=0.9\linewidth]{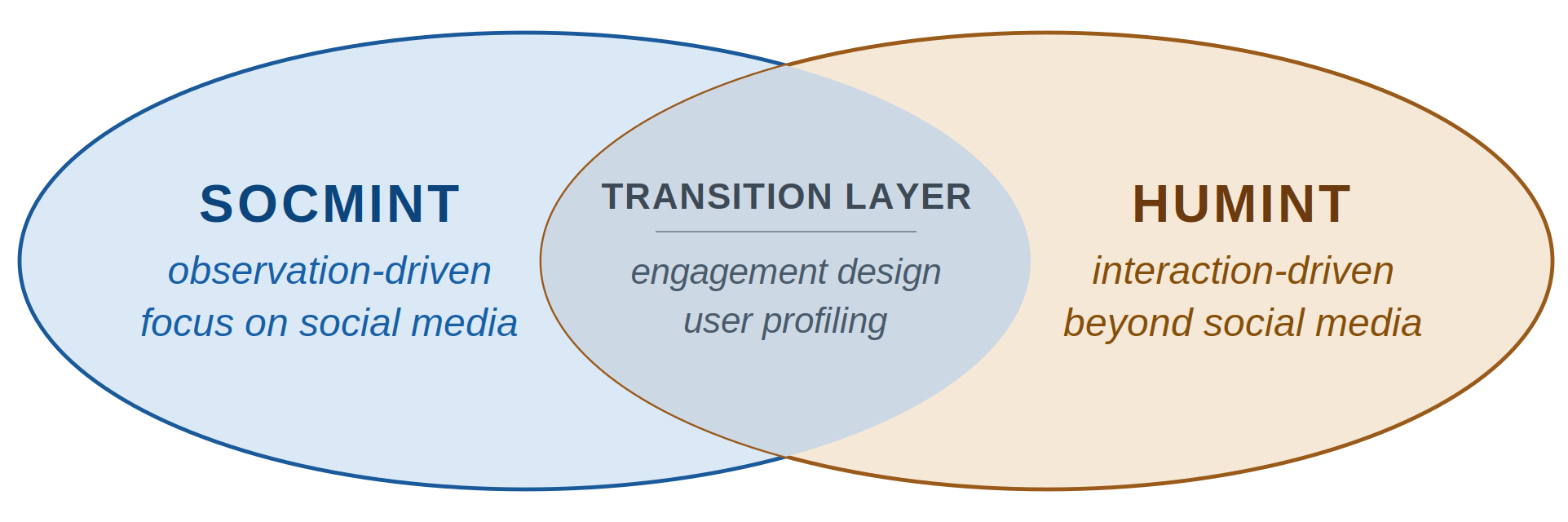}
    \caption{The SOCMINT and Cyber-HUMINT continuum: positioning of this work.}
    \label{fig:venn_paper_scope}
\end{figure}

\subsection{Major contributions of this work}

lThis work makes that exposure measurable so that individuals and organizations can assess their own risk before an adversary does. We unify the observation-to-engagement continuum into a single capability framework and operationalize its core as an LLM-augmented system, evaluated with real users (Fig.~\ref{fig:venn_paper_scope}). To our knowledge, this is the first study that measures not only how accurately such profiles can be built, but also their impact on users, specifically how far they translate into susceptibility to social engineering. Our contributions are as follows:

\begin{enumerate}
    \item \textbf{A unified capability framework} (Sec.~\ref{sec:framework}) that organizes 38 capabilities across the five phases of the Cyber Threat Intelligence (CTI) lifecycle, with profiling as the transition layer between passive collection and active engagement. We derive it from established intelligence doctrine and validate it via expert elicitation (Sec.~\ref{sec:framework_validation}; content validity, average I-CVI = 0.90; inter-rater agreement, Fleiss' $\kappa=0.92$). No single source we surveyed spans more than $\sim$68\% of these capabilities. 

    \item \textbf{An LLM-augmented multi-agent system} (Sec.~\ref{sec:llm-socmint}) that operationalizes the framework's core path, including cross-platform extraction, inference of attack-relevant latent attributes, and susceptibility assessment. In contrast to prior isolated, single-purpose tools, it orchestrates agents across platforms into one pipeline yielding traceable profiles, as analyst support rather than a replacement for established tooling. The design is informed by an analysis of the defenses deployed by 5 major social-media platforms.

    \item \textbf{A human-subjects study} (Sec.~\ref{sec:results}) measuring the core path in practice: extraction and inference accuracy on participants' own data, and the extent to which the resulting profiles anticipate the personalized engagement a social-engineering attack would exploit, marking the point where exposure becomes operational. We compare results against rule- and NER-based baselines (Sec.~\ref{subsubsec:baselinecomparison}).
\end{enumerate}

\section{Background and Related Works} \label{sec:background}

Measuring exposure along the observation-to-engage\-ment continuum requires grounding it in existing work and its gaps, and analyzing the dual-use practice around it (adversarial exploitation and the defensive disciplines that mirror it), the tools that operationalize both, and the platform constraints that bound them.

\subsection{Social-Media Profiling and Intelligence} \label{subsec:soa}
Profiling individuals from social media spans many research strands that have advanced largely in isolation, including cybersecurity, LLM-based attribute inference, extraction pipelines, and user modeling. 

In \textbf{cybersecurity}, a substantial body of work targets cyber threat intelligence (CTI). \emph{Zhao et al.}~\cite{timiner2020} automatically extract and categorize CTI from social-media streams, and \emph{Guo et al.}~\cite{tifusion2023} fuse threat signals from heterogeneous sources into structured intelligence. However, both focus on threat indicators and reports rather than the individuals behind them. From a framework perspective, \emph{Pastor-Galindo et al.}~\cite{pastor-galindo_not_2020} survey the broader OSINT landscape with a section on social media, while \emph{Teti}~\cite{teti} conceptualizes the convergence of passive observation and active engagement through a data-capture platform, but without a technical implementation. On the offensive side, \emph{Heiding et al.}~\cite{heiding2026} automate the full spear-phishing pipeline (reconnaissance, target profiling, lure generation) with click-through rates matching human experts. Focusing on lure generation, \emph{Hazell}~\cite{hazell2023} and \emph{Saha Roy et al.}~\cite{phishbots2024} produce convincing personalized phishing at scale and marginal cost, bypassing commercial guardrails. Yet the profile remains an opaque intermediate artifact, validated only by attack success. A parallel literature trains LLM and transformer classifiers to detect threats, phishing, and abusive content, which lies outside our scope. For example, \emph{San Biagio et al.}~\cite{biagio_new_2021,biagio_marple_2024} combine ML and NLP to identify threats across the surface web, social media, and darknets at up to 95\% accuracy, and \emph{Mishra et al.}~\cite{smishing2020} flag smishing from message content and URL behavior.

One line of work uses \textbf{LLMs to infer attributes} users never disclosed. \emph{Alizadeh et al.}~\cite{alizadeh_web-browsing_2025} show that web-browsing LLMs (GPT-4o, GPT-o3, Llama-3-8B-Web) infer demographic attributes from X/Twitter usernames, drawing on bios and follow networks but still guessing, with systematic bias, from minimal signals. \emph{Staab et al.}~\cite{staab_beyond_2024} show that LLMs leak sensitive attributes by inference on real Reddit profiles, and through chatbots that elicit additional cues, exposing the limits of current mitigations. \emph{Jiang and Ferrara}~\cite{jiang_social-llm_2023} and \emph{Yu}~\cite{yu_cyber_2023} combine LLM-encoded text with social-graph structure to infer partisanship and political orientation from limited footprints. These works infer specific traits (demographics, political orientation) in isolation, not within an operational profiling process spanning extraction to downstream engagement.

More broadly, in \textbf{extraction pipelines}, \emph{Arjun et al.}~\cite{arjun_anand_v_peoplexploit_2020} propose PeopleXploit, a hybrid OSINT stack orchestrating web tools for footprint profiling, yet without formal identity resolution or systematic evaluation. \emph{Wang et al.}~\cite{wang_general_2020} generate regex templates via CSS class-tag frequency analysis to extract list-structured social media content across platforms (Twitter/X, Facebook, Weibo, forum-like sites), but assume stable markup and add no semantic modeling. \emph{Sancheti et al.}~\cite{sancheti_llm_2024} use GPT-3.5 to extract 14 structured attributes (education, work experience, locations, projects, certifications) from search-engine results to disambiguate same-name individuals, but remain web-centric and decoupled from CTI.

On \textbf{user modeling and stance detection}, a large body of work models users and their opinions from social data: \emph{Tan and Jiang}~\cite{tan_user_2023} and \emph{Aldayel et al.}~\cite{aldayel2021} survey, respectively, LLM-based user modeling and stance detection, but address security only marginally, with emphasis on platform integrity (fraud, misinformation) rather than intelligence. Concrete efforts remain narrow in scope: \emph{Mbarek et al.}~\cite{mbarek2022} build a user profile by correlating multiple online social networks, but for a single clinical task and with pre-LLM methods; \emph{Rahimzadeh et al.}~\cite{rahimzadeh2025} apply an LLM stance-detection pipeline to 6M election tweets on Twitter/X, classifying opinions against predefined claims rather than extracting attributes, on a single platform and language.

Across these areas, LLM use stays confined to extraction and classification, rarely reaching latent attribute inference or engagement support, while systems remain isolated tools rather than orchestrated, cross-platform architectures. To the best of our knowledge, no prior work builds a structured, traceable profile carried from observation to engagement, nor measures the exposure it creates. Our framework addresses both gaps: it operationalizes the observation-to-engagement continuum with an LLM-augmented multi-agent system (Sec.~\ref{sec:llm-socmint}) and measures exposure through a human-subjects study (Sec.~\ref{sec:results}).

\subsection{Adversarial Use in Practice} \label{subsec:dual-use}

\begin{table*}
\centering
\caption{Documented adversarial campaigns exploiting social-media intelligence.}
\label{tab:threat-actor-mapping}
\scriptsize
\renewcommand{\arraystretch}{1.5}
\setlength{\tabcolsep}{4pt}
\begin{tabularx}{\textwidth}{@{}
  >{\raggedright\arraybackslash}p{1.5cm}|
  >{\centering\arraybackslash}p{1.5cm}
  >{\centering\arraybackslash}X
  >{\raggedright\arraybackslash}p{2.1cm}|
  >{\raggedright\arraybackslash}p{2.2cm}
  >{\raggedright\arraybackslash}p{1.9cm}
  >{\raggedright\arraybackslash}p{2.3cm}
  >{\raggedright\arraybackslash}p{2.3cm}|
  >{\raggedright\arraybackslash}p{0.5cm}
@{}}
\toprule
\textbf{Threat Actor} & \textbf{Type} & \textbf{Country} & \textbf{Campaigns} & \textbf{Social Profiling} & \textbf{Human Engage.} & \textbf{Target} & \textbf{Impact} & \textbf{Ref.} \\
\midrule
Contagious Interview &
Nation State & KP &
\textit{DEV\#POPPER} &
Profile Scraping \newline Job/Role Enum. &
Fake Job Offers &
Employees/Insiders &
Financial Theft &
\cite{mitre_contagious_interview} \\
\rowcolor{gray!10}
Lazarus Group &
Nation State & KP &
\textit{Op. Dream Job \newline Op. In(ter)ception} &
Job/Role Enum. \newline People Network &
Fake Job Offers \newline Spear Phishing &
Executives \newline Employees/Insiders &
Financial Theft \newline Data Breach &
\cite{mitre_lazarus} \\
EXOTIC LILY &
Cybercriminal & RU &
\textit{Bumblebee \newline BazarLoader} &
Profile Scraping \newline Job/Role Enum. &
Spear Phishing \newline Impersonation &
Executives \newline Employees/Insiders &
Data Breach \newline Service Disruption &
\cite{mitre_exotic_lily} \\
\rowcolor{gray!10}
Kimsuky &
Nation State & KP &
\textit{BabyShark \newline AppleSeed} &
Profile Scraping \newline Behavioral Analysis &
Impersonation &
Journalists/Activists &
Data Breach &
\cite{mitre_kimsuky} \\
APT42 &
Nation State & IR &
\textit{Charming Kitten \newline SpearSpecter} &
Profile Scraping \newline Behavioral Analysis &
Impersonation \newline Disinformation &
Political Figures &
Electoral Interf. \newline Reputation Damage &
\cite{mandiant_apt42} \\
\rowcolor{gray!10}
Scattered Spider &
Cybercriminal & US/UK &
\textit{MGM Breach \newline Caesars Breach} &
Job/Role Enum. \newline People Network &
Voice Phishing &
IT/Helpdesk &
Service Disruption \newline Data Breach &
\cite{cisa_scattered_spider} \\
UNC6040 &
Cybercriminal & -- &
\textit{Salesforce Data Theft Campaign} &
Job/Role Enum. &
Voice Phishing &
IT/Helpdesk &
Data Breach, Financial Theft &
\cite{google_unc6040} \\
\rowcolor{gray!10}
Anonymous &
Hacktivist & -- &
\textit{OpKKK, \newline HoodsOff} &
Profile Scraping \newline People Network &
Impersonation &
General Public &
Doxing &
\cite{bbc_opkkk} \\
\bottomrule
\end{tabularx}
\end{table*}

Attack campaigns confirm that social media has become an instrument of collection and engagement for offensive operations. Across eight documented campaigns (Tab.~\ref{tab:threat-actor-mapping}), attackers follow recurring patterns of social profiling and human engagement, varying in target and impact. \textit{Contagious Interview}~\cite{mitre_contagious_interview} and \textit{Lazarus}~\cite{mitre_lazarus,mitre_dream_job} pose as recruiters on LinkedIn and GitHub, luring software developers and defense staff into fake hiring processes that deliver malware. \textit{Kimsuky}~\cite{mitre_kimsuky,fbi_kimsuky} and \textit{APT42}~\cite{mandiant_apt42} instead pose as fabricated journalists and researchers, building trust with academics, policy experts, and political figures before deploying malware or harvesting credentials. \textit{EXOTIC LILY}~\cite{mitre_exotic_lily,google_exotic_lily} brokers initial access, building employee personas from LinkedIn and Crunchbase to plant ransomware; \textit{Scattered Spider}~\cite{cisa_scattered_spider} and \textit{UNC6040}~\cite{google_unc6040} use voice phishing against IT help desks and cloud administrators to hijack accounts and exfiltrate data. \textit{Anonymous}~\cite{bbc_opkkk} combines OSINT and social engineering to dox roughly 1{,}000 members of extremist groups for hacktivism purposes.     

Yet this groundwork is flattened in how intrusions are modeled: the kill chain~\cite{lockheed2011killchain} runs from reconnaissance to objectives, capturing the technical attack but compressing target profiling into a single upstream step rather than a structured phase. This pre-intrusion phase is where exposure is built and where defense must act first. 

To recover what the kill chain omits, we turn to defenders' intelligence disciplines, which already treat collection, profiling, and engagement as structured capabilities. The framework developed in this work builds on them to make this phase visible.

\subsection{SOCMINT and Cyber-HUMINT definitions} \label{subsec:definitions}
Neither social media intelligence discipline has a single universally accepted definition. However, a shared operational understanding is necessary to construct a coherent capability framework spanning both domains. 

Social Media Intelligence (SOCMINT) remains a relatively young field, and its definition continues to evolve with the platforms it studies. For the purposes of this work, we adopt the following formulation:

\noindent\rule{\linewidth}{0.4pt}

\textit{\textbf{SOCMINT} refers to intelligence derived from the collection and analysis of social media data, typically through passive, observation-driven methods, in compliance with privacy and transparency constraints. It provides near real-time situational awareness by enabling the collection and clustering of content that clarifies unfolding events, accelerates emerging threats identification, offers insight into groups, and may support the identification of criminal intent.}

\noindent\rule{\linewidth}{0.4pt}

This definition draws on established sources, each emphasizing a different facet. The \emph{National Policing Institute} focuses on social platforms as web-based systems for communication within virtual communities \cite{npi2022}; \emph{Omand et al.} \cite{omand2012socmint} place SOCMINT within the intelligence spectrum, alongside IMINT, HUMINT, and SIGINT; and the \emph{Health Sector Cybersecurity Coordination Center (HC3)} \cite{hhs2024} stresses its dual-use nature, emphasizing metadata (birth date, geolocation, social ties, workplace) alongside user-published content.

By contrast, HUMINT principles are more stable across doctrinal sources. We therefore adopt the following operational definition:

\noindent\rule{\linewidth}{0.4pt}

\textit{\textbf{Cyber-HUMINT} refers to intelligence derived from information collected by human operators and primarily provided by human sources within digital contexts. It transforms passive surveillance into active collection and is achieved through observing or directly communicating with people, according to management and governance practices specific to the source type.}

\noindent\rule{\linewidth}{0.4pt}

\emph{NATO AJP-2}, \emph{UK Joint Doctrine}, and \emph{INCIBE} stress operator training and HUMINT modalities, including debriefing, source handling, questioning, interrogation, and liaison \cite{nato2016,incibe2024}; these sources mark the boundary between passively collected surveillance outputs and their active conversion into intelligence. The \emph{Italian Security Intelligence System} highlights the need to adapt procedures to each source \cite{sisr2024}, while the \emph{CIA} adds a governance layer through formal authorization and cost--risk assessment \cite{ciaHUMINT}. \emph{U.S. Army} doctrine specifies operational-level procedures for collector employment and elicitation from sources \cite{fm2223}.

SOCMINT and Cyber-HUMINT are complementary yet distinct: the former observes, the latter engages. \emph{User profiling} bridges them, turning passively collected social media data into a structured profile that can support engagement. Yet prior work models this continuum only conceptually, without a formal capability model, leaving the continuum unmeasured. Providing such a model is therefore the objective of Sec.~\ref{sec:framework}.

\subsection{Tools for Social Media Analysis} 
\label{subsec:socmint-tools}

The security community has developed a wide range of tools that address specific aspects of identity correlation and attribute extraction \cite{pastor-galindo_not_2020}.

\textbf{Enumeration tools} verify whether a given identifier exists across multiple platforms. \textit{Sherlock} \cite{sherlock} relies on HTTP status codes, but produces many false positives. \textit{WhatsMyName} \cite{whatsmyname} improves precision through a curated dataset of platform-specific response patterns, at the cost of continuous signature maintenance. \textit{Maigret} \cite{maigret} combines both approaches, extracting basic profile attributes (name, location, bio) via DOM parsing. \textit{Holehe} \cite{holehe} leverages password-recovery endpoints to detect account registrations without triggering user notifications on supported platforms.
   
\textbf{Aggregation frameworks} integrate multiple enumeration modules and data sources into unified workflows, via modular architectures (\textit{SpiderFoot} \cite{spiderfoot}) or graph-based visual analytics (\textit{Maltego} \cite{maltego}). These require significant analyst expertise in setup and result interpretation, making them power-user tools rather than turnkey solutions.

\textbf{Commercial platforms} leverage proprietary data access and advanced analytics, extending coverage beyond publicly scrapable information. \textit{Pipl} \cite{pipl} aggregates records from people-search databases, public records, and social media APIs to build comprehensive identity profiles. \textit{PimEyes} \cite{pimeyes} enables image-based identity search via facial recognition.

Existing tools stop short of profiling itself: they generate false positives, require frequent maintenance as platforms evolve, and leave analysts to manually fuse fragmented data, infer unstated attributes, and assess relevance. Deterministic methods, such as regular expressions and named-entity recognition, handle these tasks poorly. The LLM-augmented architecture proposed in this work addresses this gap by adding semantic parsing and latent attribute inference to the pipeline, generating structured profiles while keeping human judgment central.

\begin{table*}[t]
\centering
\caption{Comparative summary of social media platform security constraints.}
\label{tab:platform-constraints}
\scriptsize
\renewcommand{\arraystretch}{1.5}
\setlength{\tabcolsep}{5pt}
\begin{tabular}{@{} l >{\raggedright}p{3.1cm} >{\raggedright}p{3.8cm} >{\raggedright}p{4.8cm} >{\raggedright\arraybackslash}p{2.5cm} @{}}
\toprule
\textbf{Platform} & \textbf{User Visibility} & \textbf{Platform / API Access} & \textbf{Abuse Resilience} & \textbf{ToS / Case Law} \\
\midrule
Facebook & User-controlled \newline (default: Friends) & Graph API; OAuth (user / app / page tokens); App Review & EDM team; rate limits; behavioral analysis; client-side defenses; account ban & \textit{Meta v. Bright Data} \\
\midrule
LinkedIn & User-controlled \newline (default: Public); connection-degree visibility & OAuth (self only); Marketing / Talent API (no scraping) & CASAL; behavioral analysis; content moderation; anti-scraping; ML real-time risk assessment; account ban & \textit{hiQ v. LinkedIn} \\
\midrule
Instagram & User-controlled \newline (default: Public) & Graph API; OAuth (Business / Creator only); App Review & EDM team; rate limits; behavioral analysis; client-side defenses; account ban & \textit{Meta v. Bright Data} \\
\midrule
X/Twitter & User-controlled \newline (default: Public) & Paid tiers (Free/Basic/Pro); OAuth (app + user level) & Rate limits; anti-spam; account ban & \textit{X v. Bright Data} \\
\midrule
TikTok & User-controlled \newline (default: Public) & Research API; Business API (no scraping); VCE (non-profit only) & CAPTCHAs; rate limits; behavioral analysis; anti-scraping & \textit{EC v. TikTok} \\
\bottomrule
\end{tabular}
\end{table*}

\subsection{Security Constraints of Social Platforms} 
\label{subsec:security-constraints}

Platform APIs have evolved from general-purpose data extraction channels into tightly controlled interfaces designed to reduce privacy leakage and automated abuse. This platform-governed access model also affects research reproducibility \cite{davidson2023platform}. We identify four layers of constraints that shape data accessibility: these constraints define the operational boundaries within which both legitimate and adversarial data collection must occur. This section examines how major platforms implement them, providing an overview in Tab.~\ref{tab:platform-constraints}. 

\textbf{User-controlled visibility.} Privacy settings place exposure under user control: content not shared stays inaccessible regardless of API capability. Yet \textit{Instagram}, \textit{X}, and \textit{TikTok} accounts are public by default \cite{instagram_privacy,x_protected_posts,tiktok_privacy_settings}. Opting into private mode restricts posts and content to approved followers, though \textit{TikTok} keeps nickname, username, and profile photo public regardless and defaults under-16 accounts to private. \textit{LinkedIn} instead grades visibility by connection degree -- full for 1st-degree, progressively reduced for 2nd-, 3rd-degree, and out-of-network users \cite{linkedin_privacy_policy}. \textit{Facebook} exposes granular per-element controls with awareness tools; \textit{X} additionally allows disabling lookup by email or phone.

\textbf{Platform-controlled authorization.} Retrievable data is the intersection of granted scopes, access tier, and endpoint policy: OAuth tokens are bound to the authenticated user, so endpoints return only self-data or aggregated signals. Higher-risk scopes require platform approval and explicit user consent. Together, these constraints block individual enumeration and large-scale social-graph reconstruction. 

\textit{Facebook}'s Graph API scopes tokens as user (individual data, with consent), app (no user data), or page (management and insights) \cite{meta_access_tokens}; since v2.0 (2014), \texttt{/me/friends} returns only friends who also authorized the app, ending bulk relationship harvesting \cite{meta_changelog_v2}, and Dev-to-Live transition requires an App Review tightened after Cambridge Analytica \cite{meta_app_review}. 

\textit{LinkedIn} is approval-gated by business line: its only self-serve API returns the authenticating user's own profile \cite{linkedin_oidc}, while Marketing and Talent APIs are role-bound to advertiser or organization-admin accounts and policy-restricted from data export or third-party combination \cite{linkedin_getting_access}. 

\textit{Instagram} retired its Basic Display API in December 2024; third-party access now requires the Graph API, limited to Business and Creator accounts (presence management, analytics, hashtag search), with endpoints gated by granular permissions and App Review \cite{instagram_api}.

\textit{TikTok} offers no general public-data API: its Research API admits only vetted academics in the US and EEA (independence, funding disclosure, ethical review), with a restricted Virtual Compute Environment for non-profits; Business API exposes no raw public data \cite{tiktok_research_api}.

\textit{X} tiers access by paid plan (Free, Basic, Pro) with rising quotas and features such as filtered streams and full-archive search; it separates App-Only tokens (public data) from User Context authentication (private or on-behalf access) \cite{x_api_access,x_api_auth}.

\textbf{Abuse resilience.} Platforms enforce access dynamically through rate limits, behavioral fingerprinting, bot detection, and CAPTCHAs, and resist scraping through DOM randomization, virtualization, and lazy loading. They adapt in real time to throttle, challenge, or ban suspicious activity. 

\emph{Facebook} runs a dedicated External Data Misuse team with rate and data limits plus behavioral analysis \cite{meta_combat_scraping_2021}. Our analysis (Sec.~\ref{sec:llm-socmint}) identifies further client-side defenses, including randomized CSS class names (which force reliance on stable ARIA attributes), lazy loading with placeholders, text fragmented across \texttt{<span>} tags, and DOM virtualization that drops off-viewport posts while scrolling. Together, these mechanisms defeat reliable single-pass bulk extraction. \emph{Instagram} inherits Meta's centralized detection and rate limiting, restricting accounts that scrape \cite{instagram_scraping_restricted}. \emph{LinkedIn} centers its Trust ML stack on CASAL (Centralized Abuse Scoring As a Library), which scores abuse (fake accounts, account takeover, moderation, scraping) via retrained models \cite{linkedin_casal_2023}.

\emph{X} rate-limits per application and user, locks accounts behind anti-spam challenges on suspicion, and permanently suspends non-API automation or rate-limit bypass \cite{x_rate_limits,x_automation_rules}. \emph{TikTok} combines CAPTCHAs, rate limiting, and behavioral analysis of devices, networks, and interaction patterns, plus undisclosed controls \cite{tiktok_combat_scraping}.

\textbf{Terms of Service.} Platform Terms uniformly prohibit scraping, fake accounts, and circumvention of technical controls, exposing violators to account termination and legal action \cite{meta_tos, linkedin_user_agreement, x_tos, tiktok_tos}. Yet courts have repeatedly held that scraping public, logged-out data does not breach these Terms, since non-authenticated users never assented to them and data lack authentication barriers: \emph{hiQ Labs v. LinkedIn} (9th Cir. 2022) \cite{hiq_v_linkedin_2022}, \emph{Meta v. Bright Data} \cite{meta_v_bright_data_2024}, \emph{X Corp. v. Bright Data} (N.D. Cal. 2024) \cite{x_v_bright_data_2024}. The European Commission's 2025 preliminary finding also faults \textit{TikTok} for over-restricting researcher access under the Digital Services Act \cite{ec_tiktok_dsa_2025}.

Across the five platforms, defensive postures converge (rate limiting, behavioral analysis, ToS restrictions), but API availability diverges. This asymmetry between platform-controlled access and research needs motivates alternative collection. Accordingly, our study collected only data participants had made publicly visible, accessed through the authenticated sessions that platform controls require.

\section{Methodology}
\label{sec:methodology}

Our research followed a six-phase design that spans doctrinal foundations to operational validation, grounded in established methodologies \cite{nickerson,braun2006,polit2006,landis1977}.

\textbf{Phase 1 -- Domain foundation.} We anchored the study in intelligence doctrine and the post-API access landscape, defining the SOCMINT--Cyber-HUMINT disciplines and their collection constraints (Sec.~\ref{sec:background}).

\textbf{Phase 2 -- Framework derivation.} We developed the capability taxonomy in line with established taxonomy-development guidance, following an empirical\hyp{}to\hyp{}conceptual strategy~\cite{nickerson}. We derived 38 capabilities from eight doctrinal and operational sources~\cite{nato2016,sisr2024,incibe2024,ciaHUMINT,npi2022,hhs2024,omand2012socmint,fm2223} through \textit{Braun and Clarke} thematic analysis~\cite{braun2006}: we extracted operational concepts near-verbatim from each source, grouped them into thematic clusters aligned with the five CTI lifecycle phases, and refined them into discrete capabilities, each with explicit traceability to its originating passages (Sec.~\ref{sec:framework}).

\textbf{Phase 3 -- Framework validation.} Following the content\hyp{}validation practice of \emph{Polit and Beck}~\cite{polit2006}, we submitted the framework to a panel of 13 cybersecurity professionals recruited through targeted invitations (Sec.~\ref{sec:framework_validation}). The panel covered three functional areas: Cyber Threat Intelligence (6), Cyber Defense (4), and Strategy and Risk (3). Participants were based in three countries and had 2 to 13 years of experience. After a briefing session, experts received the complete taxonomy and were asked to assess:
\begin{enumerate}[leftmargin=*,nosep,label=\alph*)]
\item relevance to social intelligence operations (1--5 scale), via the Item-level Content Validity Index (I-CVI) computed as the proportion of experts rating the capability $\geq 4$ (content-validity threshold: I-CVI $\geq 0.78$);
\item correctness of the assigned CTI phase, assessed via Fleiss' $\kappa$ for inter-rater agreement, with $\kappa \geq 0.61$ indicating substantial agreement~\cite{landis1977};
\item the overall framework structure (1--5 Likert), plus any superfluous or missing capabilities.
\end{enumerate}

\textbf{Phase 4 -- System architecture.} From the taxonomy, we selected the capabilities amenable to LLM automation: cross-platform extraction, latent-attribute inference, and susceptibility assessment (Sec.~\ref{sec:llm-augmentation}). We instantiated these capabilities in an LLM-augmented multi-agent pipeline, yielding traceable profiles and separating extracted facts from inferred attributes (Sec.~\ref{sec:llm-socmint}).

\textbf{Phase 5 -- Empirical validation.} We evaluated the pipe\-line in an observational human-subjects study with 14 cybersecurity professionals recruited by convenience sampling from the research team's network (Sec.~\ref{sec:results}). Inclusion required at least one active public \textit{Facebook} and \textit{LinkedIn} profile and English proficiency. Participation was voluntary, uncompensated, and outside any supervisory relationship. For each participant, the pipeline scraped publicly accessible fields, mapped them to canonical fields, and generated a profile report (Sec.~\ref{sec:architecture}). Participants then reviewed their report and completed a four-layer assessment instrument (Sec.~\ref{subsec:reporting}), which served as the authoritative ground truth for their own public content. Sessions were self-paced and asynchronous. We measured extraction performance against rule-based and NER-based baselines, inference agreement against participant self-report, and engagement susceptibility with Kendall's $\tau$. The study received ethics approval (see \nameref{sec:ethics}).

\textbf{Phase 6 -- Discussion.} We interpreted the results against prior work, drew their defensive and dual-use implications, and set out limitations and future directions (Sec.~\ref{sec:discussion}).

\section{The Social-HUMINT Framework} \label{sec:framework}

\definecolor{phbg}{RGB}{58,58,58}
\definecolor{socmint}{RGB}{26,107,191}
\definecolor{profiling}{RGB}{27,138,90}
\definecolor{humint}{RGB}{196,122,10}
\definecolor{citecolor}{RGB}{130,130,130}

\newcommand{\dotS}{\tikz[baseline=-0.5ex]\draw[fill=socmint,draw=none] circle (1.8pt);}
\newcommand{\dotP}{\tikz[baseline=-0.5ex]\draw[fill=profiling,draw=none] circle (1.8pt);}
\newcommand{\dotH}{\tikz[baseline=-0.5ex]\draw[fill=humint,draw=none] circle (1.8pt);}
\newcommand{\dotX}{\tikz[baseline=-0.5ex]\draw[fill=none,draw=gray!35,line width=0.4pt] circle (1.8pt);}

\newcommand{\SPH}{\dotS\,\dotP\,\dotH}
\newcommand{\SxH}{\dotS\,\dotX\,\dotH}
\newcommand{\Sxx}{\dotS\,\dotX\,\dotX}
\newcommand{\xxH}{\dotX\,\dotX\,\dotH}
\newcommand{\xPH}{\dotX\,\dotP\,\dotH}
\newcommand{\xPx}{\dotX\,\dotP\,\dotX}
\newcommand{\SPx}{\dotS\,\dotP\,\dotX}

\newcommand{\E}[3]{%
  \parbox[t]{\linewidth}{%
    #1\hfill#3\par
    \vspace{1pt}%
    {\scriptsize\textcolor{citecolor}{\protect#2}}%
  }%
}

\newcommand{\Eg}[3]{%
  \parbox[t]{\linewidth}{%
    #1\hfill#3\par
    \vspace{1pt}%
    {\scriptsize\textcolor{citecolor}{\protect#2}}~{\footnotesize\faCog}%
  }%
}

\begin{figure*}[t]
\centering
\caption{Social-HUMINT framework: capability classification across CTI phases.}
\label{fig:sochumint-framework}
\scriptsize
\renewcommand{\arraystretch}{1.7}
\setlength{\tabcolsep}{5pt}
\begin{tabular*}{\textwidth}{@{\extracolsep{\fill}}
  >{\raggedright}p{0.185\textwidth}
  >{\raggedright}p{0.185\textwidth}
  >{\raggedright}p{0.185\textwidth}
  >{\raggedright}p{0.185\textwidth}
  >{\raggedright\arraybackslash}p{0.185\textwidth}
@{}}
\rowcolor{phbg}
\textcolor{white}{\textbf{Planning \& Direction}} &
\textcolor{white}{\textbf{Collection}} &
\textcolor{white}{\textbf{Processing}} &
\textcolor{white}{\textbf{Analysis \& Correlation}} &
\textcolor{white}{\textbf{Dissemination}} \\[2pt]

\E{Strategy definition}      {\cite{nato2016,sisr2024,ciaHUMINT,hhs2024,pastor-galindo_not_2020,fm2223}}{\SxH} &
\Eg{Passive observation}     {\cite{omand2012socmint,npi2022,hhs2024,pastor-galindo_not_2020}}{\Sxx} &
\E{Data triage \& filtering} {\cite{omand2012socmint,nato2016,incibe2024,pastor-galindo_not_2020,fm2223}}{\SxH} &
\Eg{Behavior \& pattern}     {\cite{omand2012socmint,pastor-galindo_not_2020,teti,fm2223}}{\SPH} &
\Eg{Info sharing}            {\cite{omand2012socmint,nato2016,npi2022,fm2223}}{\SxH} \\

\E{Risk assessment}          {\cite{omand2012socmint,nato2016,npi2022,fm2223}}{\SxH} &
\Eg{Automated acquisition}   {\cite{omand2012socmint,pastor-galindo_not_2020,teti}}{\SxH} &
\Eg{Multi-channel normaliz.} {\cite{nato2016,omand2012socmint,pastor-galindo_not_2020,teti}}{\SPH} &
\E{Media analysis}           {\cite{omand2012socmint,pastor-galindo_not_2020,teti}}{\SPx} &
\E{Threat reporting}         {\cite{nato2016,npi2022,fm2223}}{\SxH} \\

\E{Intel.\ requirements}     {\cite{nato2016,incibe2024,ciaHUMINT,fm2223}}{\SxH} &
\Eg{Active engagement}       {\cite{nato2016,sisr2024,incibe2024,hhs2024,teti,fm2223}}{\xxH} &
\E{KB ingestion}             {\cite{fm2223,nato2016,pastor-galindo_not_2020,teti}}{\SxH} &
\E{People network}           {\cite{omand2012socmint,nato2016,pastor-galindo_not_2020,teti,fm2223}}{\SPH} &
\E{Lessons learnt}           {\cite{nato2016,npi2022,fm2223}}{\SxH} \\

\E{Legal \& ethical}         {\cite{omand2012socmint,nato2016,npi2022,pastor-galindo_not_2020,fm2223}}{\SxH} &
\E{Multimedia collection}    {\cite{incibe2024,hhs2024,pastor-galindo_not_2020,teti}}{\SxH} &
\Eg{Entity extraction}       {\cite{omand2012socmint,nato2016,pastor-galindo_not_2020,teti}}{\SPH} &
\Eg{Trend \& temporal}       {\cite{omand2012socmint,pastor-galindo_not_2020,teti,fm2223}}{\SPH} &
\Eg{Dashboard \& visual.}    {\cite{omand2012socmint,npi2022,pastor-galindo_not_2020,teti}}{\SxH} \\

\E{Collection strategy}      {\cite{nato2016,sisr2024,incibe2024,ciaHUMINT,hhs2024,pastor-galindo_not_2020,fm2223}}{\SxH} &
\E{Crowd-sourced collect.}   {\cite{omand2012socmint,pastor-galindo_not_2020}}{\SxH} &
\E{Identity resolution}      {\cite{omand2012socmint,nato2016,pastor-galindo_not_2020}}{\SPH} &
\E{Intent \& deception}      {\cite{omand2012socmint,nato2016,npi2022,pastor-galindo_not_2020,fm2223}}{\SPH} &
\E{Intel.\ liaison}          {\cite{nato2016,ciaHUMINT,npi2022,fm2223}}{\SxH} \\

\E{Digital identity strategy} {\cite{incibe2024,npi2022,hhs2024,teti}}{\SxH} &
\E{Historical research}      {\cite{npi2022,omand2012socmint,pastor-galindo_not_2020}}{\Sxx} &
\Eg{Data enrichment}         {\cite{hhs2024,pastor-galindo_not_2020,fm2223}}{\SPH} &
\E{Debriefing}               {\cite{nato2016,fm2223}}{\xxH} &
\E{Oper.\ actionability}     {\cite{omand2012socmint,nato2016,npi2022,fm2223}}{\SxH} \\

\Eg{Tool setup}              {\cite{incibe2024,hhs2024,pastor-galindo_not_2020,teti}}{\SxH} &
&
\E{Metadata extraction}      {\cite{hhs2024,nato2016,pastor-galindo_not_2020,teti}}{\SPx} &
\Eg{Attribute inference}     {\cite{omand2012socmint,nato2016,pastor-galindo_not_2020,teti}}{\SPH} &
\\

\E{Engagement planning}      {\cite{npi2022,teti,fm2223}}{\xxH} &
&
&
\Eg{Profile construction}    {\cite{nato2016,omand2012socmint,pastor-galindo_not_2020,teti}}{\xPx} &
\\

& & &
\E{Group dynamics}           {\cite{omand2012socmint,nato2016,teti,fm2223}}{\SxH} &
\\

& & &
\E{Event \& threat detect.}  {\cite{omand2012socmint,nato2016,npi2022,pastor-galindo_not_2020,fm2223}}{\Sxx} &
\\

& & &
\E{Pivot discovery}          {\cite{fm2223,nato2016,pastor-galindo_not_2020}}{\SxH} &
\\[2pt]

\end{tabular*}

\vspace{1.0em}
{\centering\scriptsize
Discipline:\quad \dotS~SOCMINT \quad \dotP~USER PROFILING \quad \dotH~CYBER-HUMINT \qquad \faCog~\textit{LLM-augmented in our architecture}\par}

\end{figure*}

The Social-HUMINT framework formalizes 38 capabilities across the five CTI lifecycle phases, grounded in the definitions of Sec.~\ref{sec:background}~\cite{omand2012socmint,nato2016,fm2223,ciaHUMINT,pastor-galindo_not_2020,teti,incibe2024,hhs2024,npi2022,sisr2024} and organized by discipline applicability: SOCMINT (passive observation), user profiling (the data-to-profile bridge), and Cyber-HUMINT (active engagement). The passive-to-active distinction guides method selection across the continuum. The process is cyclical: disseminated intelligence informs subsequent planning.

The disciplines are not disjoint. They share a common capability set, and most capabilities recur across three interaction levels that differ only in intensity (reachable sources and degree of direct target engagement), which is precisely what a unified taxonomy should capture. \textit{Low} corresponds to passive monitoring of publicly accessible content and requires no account or contact. \textit{Medium} adds acquisition from access-controlled or subscription-based sources and low-engagement actions; this level remains in the passive-to-controlled range and spans both SOCMINT and user profiling. \textit{High} requires active engagement with individuals or communities, including messaging, posting, and information exchange, and is therefore exclusive to Cyber-HUMINT. Because traceability and operational exposure increase with intensity, the level determines the applicable legal and security constraints.

The procedural formalization below is the operational foundation for the LLM-augmented architecture of Sec.~\ref{sec:llm-socmint}. Fig.~\ref{fig:sochumint-framework} provides per-capability source traceability and marks with a gear (\faCog) the capabilities augmented in our architecture, whose augmentation is detailed in Sec.~\ref{sec:llm-augmentation}.

\subsection{\textbf{Planning and Direction}} Establishes operational foundations before any other activity, defining strategic objectives, risk boundaries, legal constraints, and technical requirements. 

\begin{itemize}[leftmargin=*,nosep]
\item \textbf{Strategy Definition \& Prioritization.} Determines focus areas, target categories, and resource allocation.
    
\item \textbf{Risk Assessment.} Evaluates operational exposure to ensure method intrusiveness is proportionate to objectives.
    
\item \textbf{Intelligence Requirements.} Specifies information needs, granularity levels, and decision-support objectives.
    
\item \textbf{Legal \& Ethical Constraints.} Establishes regulatory boundaries and obtains required authorizations through defined approval chains and oversight mechanisms.
    
\item \textbf{Source \& Collection Strategy.} Classifies and rates intelligence sources (public, authenticated, paid) by reliability, defining collection modes, sampling strategies and continuous maintenance.
    
\item \textbf{Digital Identity Strategy.} Creates and maintains operational personas, from access aliases for passive observation to full cover identities (backstory and justification) for sustained engagement.
    
\item \textbf{Tool Setup.} Configures the technical infrastructure for downstream collection, inference, and engagement, including data acquisition pipelines, LLM engines, and interaction-support systems.
    
\item \textbf{Engagement Planning.} Designs target-interaction workflows from structured profile outputs by specifying objectives, timing, channel selection, messaging strategy, and contingency protocols.
\end{itemize}

\subsection{\textbf{Collection}} Acquires raw evidence through passive and active methods across social platforms, human interactions, and distributed sources.

\begin{itemize}[leftmargin=*,nosep]
    \item \textbf{Passive Observation.} Monitors targets without direct interaction, including feed tracking, public-activity surveillance, and content consumption. It also includes manual reconnaissance and monitoring of closed groups.
    
    \item \textbf{Automated Data Acquisition.} Executes programmatic collection through API connectors, scrapers, streaming endpoints, or LLM-powered bots. Enables bulk profile extraction for downstream inference pipelines.
    
    \item \textbf{Active Engagement Execution.} Conducts direct target interaction based on predefined engagement plans, including tactical questioning, interrogation, elicitation, social engineering, group participation, and platform actions (posting, commenting, reacting). Conversational probing validates or refines inferred profile attributes.
    
    \item \textbf{Multimedia Content Collection.} Acquires target\hyp{}generated content across formats (text posts, comments, images, video, audio, shared media), including content captured during active engagement. Preserves original format for downstream content analysis.
    
    \item \textbf{Crowd-sourced Collection.} Gathers non-targeted signals from distributed sources, such as anonymous tips, community reports, and citizen journalism during unfolding events.
    
    \item \textbf{Historical Research.} Retrieves archived target data from web archives, cached content, deleted-post recovery, and legacy platform records. This capability reconstructs past digital footprints for longitudinal analysis and evidence preservation.
    
\end{itemize}

\subsection{\textbf{Processing}} Transforms raw collected data into structured, queryable intelligence by normalizing heterogeneous sources, resolving identities, and extracting entities.

\begin{itemize}[leftmargin=*,nosep]
    \item \textbf{Data Triage \& Filtering.} Filters noise, duplicates, and irrelevant content; grades source reliability and quality to prioritize high-value data. Validation may draw on active interaction.
    
    \item \textbf{Multi-channel Normalization.} Standardizes data from heterogeneous sources and channels into a unified schema. Harmonizes fields, dates, relationships, and structures, including multilingual translation.
    
    \item \textbf{Knowledge Base Ingestion.} Loads collected data into structured repositories that enable query, retrieval, and cross-referencing. Establishes data lineage and maintains source attribution throughout the pipeline.
    
    \item \textbf{Entity Extraction.} Extracts explicit entities from structured/unstructured content (names, locations, orgs, dates, events, relationships), via NER and pattern matching.

    \item \textbf{Identity Resolution.} Links accounts, aliases, and cross-platform identities to single individuals via metadata and naming patterns, establishing attribution confidence.
    
    \item \textbf{Data Enrichment.} Augments social media data with external sources (public records, OSINT databases, corporate registries, offline-collected information) to fill gaps not addressable by social-platform collection alone.

    \item \textbf{Metadata Extraction.} Captures technical attributes embedded in collected content, including geolocation, timestamps, device fingerprints, and platform-specific identifiers.
    
\end{itemize}

\subsection{\textbf{Analysis \& Correlation}} Transforms processed intelligence into structured insights and conclusions that inform dissemination and CTI operations. This phase interprets data, establishes causal relationships, and determines contextual significance.

\begin{itemize}[leftmargin=*,nosep]
    \item \textbf{Behavioral \& Pattern Analysis.} Detects anomalies, routine deviations, and significant changes in target activity through behavior and pattern analysis over time. 

    \item \textbf{Media Content Analysis.} Performs semantic analysis of multimedia content (images, video, audio), extracting sentiment, objects, locations, or contextual meaning.

    \item \textbf{People Network Analysis.} Maps relationships and social structures around the target, identifying key contacts, affiliations, group memberships, or interpersonal dynamics.

    \item \textbf{Trend \& Temporal Analysis.} Analyzes the temporal evolution of activity. Detects peaks, seasonality, trend shifts, and time-based correlations.

    \item \textbf{Intent \& Deception Analysis.} Evaluates target objectives, motivations, and truthfulness, detecting deliberate misinformation and gauging the target's awareness of observation and possible countermeasures.

    \item \textbf{Debriefing.} Systematically reviews information gathered through direct engagement by analyzing target responses, evaluating reliability, and reassessing prior classifications when new evidence contradicts them.

    \item \textbf{Attribute Inference.} Derives non-explicit target attrib\-utes through analytical reasoning and LLM-augmented processing (e.g., age ranges, interests, affiliations, behavioral traits). Distinguishes inferred attributes from extracted facts with associated confidence levels.

    \item \textbf{Profile Construction.} Assembles extracted entities and inferred attributes into standardized target profiles, preserving the fact/inference distinction.
    
    \item \textbf{Event \& Threat Detection.} Monitors data streams to identify emerging events, hostile actors, coordinated campaigns, and risk indicators requiring immediate attention.
    
    \item \textbf{Group Dynamics Analysis.} Examines collective behavior within target groups, including internal hierarchies, influence distribution, leadership roles, subgroup formation, and indicators of cohesion or fragmentation.

    \item \textbf{Pivot Discovery.} Identifies new targets, leads, and intelligence opportunities emerging from current analysis for subsequent collection cycles.
\end{itemize}

\subsection{\textbf{Dissemination}} Delivers intelligence products to authorized stakeholders and transfers knowledge across organizational boundaries.

\begin{itemize}[leftmargin=*,nosep]
    \item \textbf{Info Sharing.} Distributes validated intelligence to authorized stakeholders through secure channels, with timely delivery based on clearance and operational needs.

    \item \textbf{Threat Reporting.} Produces structured reports on identified threats, hostile actors, and ongoing campaigns, including context and recommended response priorities.

    \item \textbf{Recommendations \& Lessons Learned.} Documents operational insights and methodological improvements, systematically incorporating findings into subsequent planning and collection cycles.

    \item \textbf{Dashboard \& Dynamic Visualization.} Provides visual interfaces for monitoring and presenting outputs, enabling decision-makers to interpret complex data rapidly.

    \item \textbf{Intelligence Liaison.} Establishes formal coordination and structured intelligence exchange with external agencies, operational partners, and allied organizations.

    \item \textbf{Operational Actionability.} Translates validated intelligence into actionable guidance on how to interpret and use it in context, including response protocols and confidence assessment before operational use.
  
\end{itemize}

\subsection{LLM Augmentation across CTI Phases} \label{sec:llm-augmentation}

Most framework capabilities involve unstructured-text interpretation, cross-source reasoning, or narrative generation. These tasks are suitable for LLM augmentation across the CTI cycle. This section maps that potential phase by phase, separating opportunities from the capabilities our case study evaluates empirically (Sec.~\ref{sec:llm-socmint}).

\textit{\textbf{Planning \& Direction.}} LLMs can inform strategy definition from open-source synthesis, map regulatory constraints against planned methods (GDPR, platform ToS, jurisdiction), and derive intelligence requirements from threat models. Digital identity strategy (credible personas with synthetic content and automated response logic) and engagement planning (profile-informed narrative construction) are natural fits for generative models. In our pipeline planning stays analyst-driven, but tool setup is implemented directly: the analyst configures the Langflow pipeline, with LLM agents as its core reasoning engine.

\textit{\textbf{Collection.}} For passive observation, LLMs summarize monitored feeds, detect signals in high-volume streams, and extract digital footprints from archived or cached sources. Active engagement benefits from content generation, from crafting contextually appropriate posts and comments to fully automated bot-driven interaction through the personas established in planning. Multimodal models extend collection to images and video by extracting contextual signals (e.g., geolocation) without dedicated computer-vision methods. Our pipeline operationalizes automated data acquisition: the SOCMINT stage scrapes Facebook and LinkedIn, and LLM agents parse all raw page content. Active engagement is evaluated indirectly through the Social Engineering assessment (Sec.~\ref{subsubsec:results-quiz}).

\textit{\textbf{Processing.}} LLMs filter noise during parsing, harmonize heterogeneous sources into unified schemas, and resolve identities across platforms, complementing regex extraction with context-aware parsing; knowledge-base ingestion is integral, feeding both extraction and inference. Entity extraction, attribute inference, and profile construction form the core of our implementation: the Profiling stage turns raw scraped data into typed fields and higher-order inferences that would otherwise need analyst judgment (Sec.~\ref{subsec:profile-model}), and the Profiler Agent fuses LinkedIn and Facebook outputs through conflict-resolution rules favoring each source's domain of strength.

\textit{\textbf{Analysis \& Correlation.}} Reasoning across multiple profile entries, LLMs support behavioral and pattern analysis, intent assessment, and trend detection; multimodal models add semantic interpretation of media; people-network analysis, event and threat detection, assisted debriefing, and pivot discovery are further augmentation surfaces. Our pipeline demonstrates behavioral analysis through inferred attributes (e.g., communication style, career trajectory, privacy awareness), extending it with a personalized assessment of whether the collected data supports credible social engineering.

\textit{\textbf{Dissemination.}} LLMs can generate threat reports, liaison summaries, dashboard narratives, and recommendations from CTI outputs, adapting format and depth to audiences from operational briefings to executive summaries. Our architecture produces a structured intelligence dossier, while natural-language reporting is a direct extension.

\section{Case Study: LLM-Augmented Social-HUMINT} \label{sec:llm-socmint}

This section operationalizes a core subset of the Social-HUMINT framework (Sec.~\ref{sec:framework}) through a Langflow\hyp{}orchestrated multi-agent system, showing how LLM augmentation supports user profiling across the Collection\hyp{}Processing\hyp{}Analysis chain. The implementation covers automated cross-platform extraction, latent attribute inference, and assessment of whether collected profile data can support credible social-engineering operations.

\begin{figure*}[t]
    \centering
    \includegraphics[width=1\linewidth]{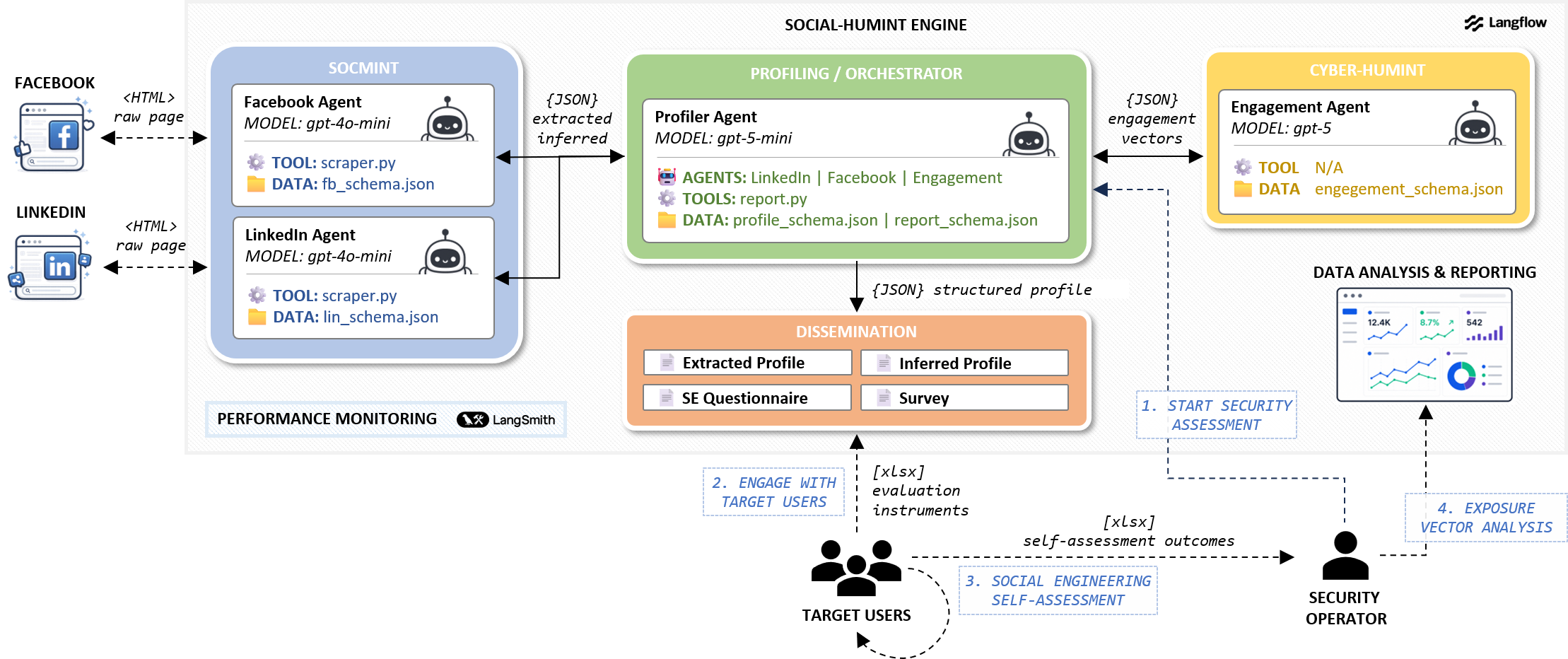}
    \caption{LLM-augmented Social-HUMINT pipeline architecture.} 
    \label{fig:langflow_architecture}
\end{figure*}

\subsection{System Architecture} 
\label{sec:architecture}

The architecture integrates three core components: a scraping layer for social-media data extraction, a multi-agent orchestration system for intelligence processing, and an observability infrastructure for auditability and reproducibility. We orchestrate the pipeline with Langflow~\cite{langflow2024} and trace execution with LangSmith, which records the full call graph and key operational metrics (latency, tokens, and cost).

User profiling serves as the bridge between observation-driven and interaction-driven intelligence practices (Fig.~\ref{fig:langflow_architecture}). The \textbf{SOCMINT stage} performs passive, observation-driven collection from LinkedIn and Facebook through platform-specific scrapers. The \textbf{Profiling stage} acts as a transition layer: it normalizes cross-platform evidence, infers latent attributes, and constructs structured user profiles. The \textbf{Cyber-HUMINT stage} consumes these profiles to generate targeted engagement artifacts and produce \textbf{reporting} outputs.

This modular design aligns with operational CTI workflows and accommodates the inherent volatility of social media platforms through decoupled, independently maintainable components.

\subsubsection{Scraping Strategy} 
\label{subsec:scraping}

The scraping layer follows the \textit{``dumb scraper, smart LLM''} paradigm: extraction logic is deliberately minimal, limited to raw page retrieval and lazy content loading, while all semantic parsing is delegated to downstream agents. This design improves resilience to platform-specific DOM changes, a common failure mode of regex-based or tightly coupled parsers, and reduces maintenance overhead to cookie refresh cycles.

We target LinkedIn and Facebook, which are primary OSINT sources for a target's professional and personal footprint, respectively~\cite{pastor-galindo_not_2020}. Both scrapers are custom Langflow components built on Playwright~\cite{microsoft2020playwright}, which traverses complex DOM structures and dynamic content that static HTTP requests cannot reach. Since neither platform exposes a public profile API, the scrapers authenticate by injecting existing session cookies, replicating the view available to any logged-in user, and apply standard anti-detection measures (fingerprint masking, randomized delays, and human-like lazy scrolling). Given a name rather than a URL, the scraper resolves the target through keyword search by leveraging the platform's proximity-based ranking. Extraction depth and post volume are configurable; in our runs, we used three scroll iterations to trigger lazy-loaded content and captured up to ten recent posts, maximizing downstream signal within context-window limits.

\subsubsection{Agentic Reasoning} 
\label{subsec:agentic}

The pipeline employs a multi-agent architecture in which each agent has a single responsibility. Prompts combine standard techniques (system and role prompting, task decomposition, and constraint rules)~\cite{sahoo2024prompt}, and agents follow the ReAct paradigm~\cite{yao2023react}, interleaving reasoning with tool calls. Prompt details are provided in the Supplementary Material. All outputs conform to JSON schemas with mandatory confidence scores and reasoning chains for inferred attributes, supporting auditability and human-in-the-loop validation. The pipeline comprises the following agents:

\begin{itemize}[leftmargin=*]
\item \textbf{(SOCMINT) LinkedIn Agent / Facebook Agent.} These agents consume raw scraper output and process it in two phases: extraction of explicit fields and inference of latent attributes (per the data model in Sec.~\ref{subsec:profile-model}). Each inference includes per-attribute confidence and supporting evidence. Each agent also generates three conversational topics as engagement hooks for downstream use.

\item \textbf{(Transition Layer) Profiler Agent.} Orchestrates cross-platform data fusion from LinkedIn and Facebook agents, merging outputs into a unified profile using conflict-resolution rules (LinkedIn preferred for professional attributes, Facebook for personal data). The agent merges conversational topics from both sources (up to six) and generates a balanced short bio integrating professional and personal dimensions.

\item \textbf{(Cyber-HUMINT) Engagement Agent.} This agent emulates the HUMINT analyst, drawing on the user profile to engage the target. Rather than executing engagement directly, our study uses a 12-question instrument across three social-engineering levers, based on behavioral scenarios instead of direct assessment (Sec.~\ref{subsec:reporting}).

\end{itemize}

Selected models scale with task complexity, balancing reasoning depth against operational cost:

\begin{itemize}[nosep]
\item \texttt{gpt-4o-mini-2024-07-18} for extraction;
\item \texttt{gpt-5-mini-2025-08-07} for profiling;
\item \texttt{gpt-5-2025-08-07} for active engagement.
\end{itemize}

\begin{figure*}
    \centering
    \includegraphics[width=1\linewidth]{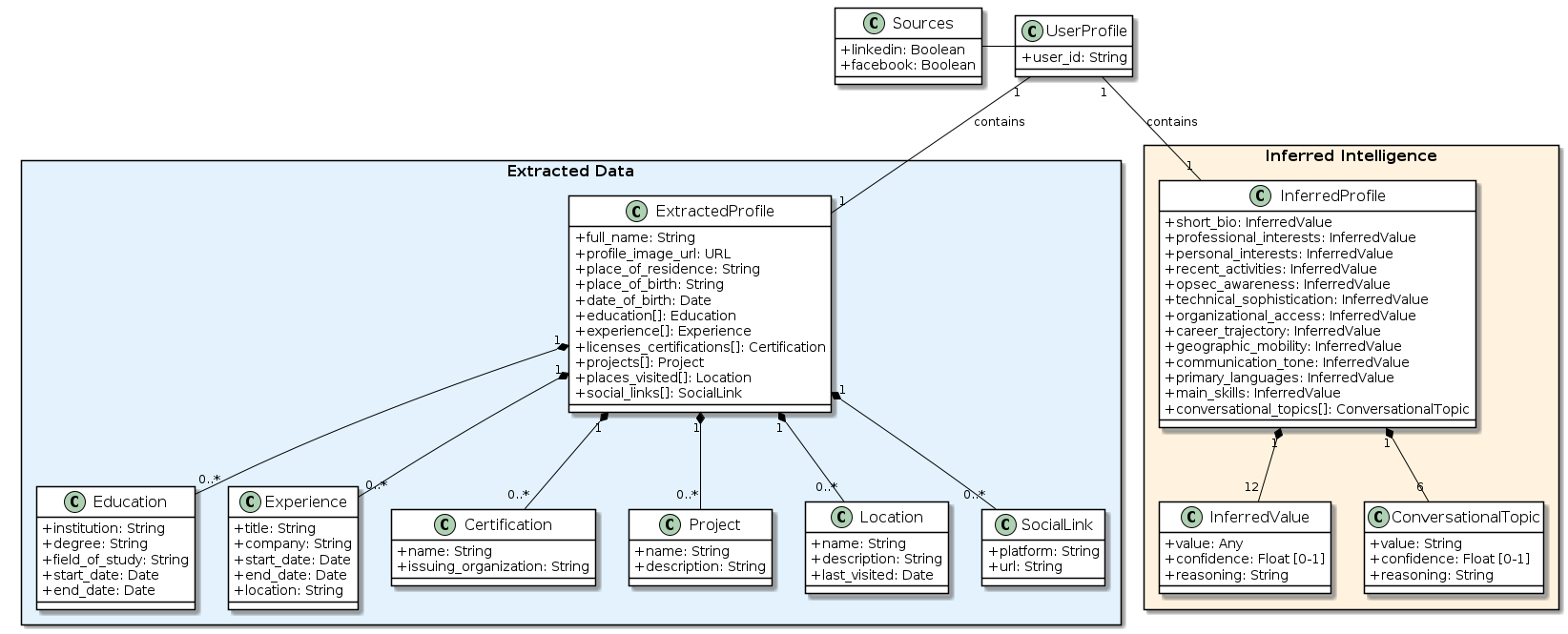}
    \caption{Two-tier user profiling data model.}
    \label{fig:user_profile_UML}
\end{figure*}

\subsubsection{User Profile Data Model} \label{subsec:profile-model}

Pipeline output is structured as a two-tier data model that formalizes the distinction between directly observable attributes and intelligence derived through LLM-based reasoning (Fig.~\ref{fig:user_profile_UML}).

The \textit{Extracted Data} layer comprises structured fields harvested from social media platforms (LinkedIn, Facebook) via automated scraping. These ground-truth observations are organized into seven normalized categories: \emph{Identity} (full name, profile image, place and date of birth, residence location), \emph{Experience} (professional positions), \emph{Education} (degrees and institutions), \emph{Certifications} (professional licenses and credentials), \emph{Projects} (published projects and portfolios), \emph{Places Visited} (geographic locations mentioned or tagged), and \emph{Social Links} (linked social media accounts).

The \textit{Inferred Intelligence} layer captures derived attributes not explicitly declared in source profiles. Each inference is encapsulated in an \textit{InferredValue} structure containing the inferred value, a confidence score (in the [0,5] range), and an explicit reasoning explanation. Inferred attributes span four functional categories: (i)~\textit{behavioral profiling}, including communication patterns, language preferences, career trajectory, and geographic mobility; (ii)~\textit{interest modeling}, which defines potential engagement vectors in professional and personal domains; (iii)~\textit{target assessment}, covering privacy awareness, technical capabilities, and organizational access level; and (iv)~\textit{engagement support}, including conversational topics derived from profile data for rapport building. A \textit{Metadata} layer tracks data provenance for source attribution and multi-platform validation.

\subsubsection{Evaluation Instruments} \label{subsec:reporting}
We evaluate the pipeline through participant self-report. Only the target can authoritatively judge whether inferred attributes are correct and whether the profile is exploitable for social engineering~\cite{goffman1959}. This mirrors intelligence practice, where hypotheses are refined through human judgment under uncertainty~\cite{lowenthal2020,heuer1999}. Four instruments operationalize this approach and enable evaluation across quantitative and qualitative metrics.
\begin{itemize}[leftmargin=*]

\item \textbf{Extracted data.} For each extracted field (Sec.~\ref{subsec:profile-model}), participants mark it \textit{correct}, \textit{incorrect}, or \textit{missing}, mapping respectively to true positives, false positives, and false negatives. True negatives are undefined (the field universe is open), so we report precision, recall, and $F_1$ rather than accuracy. Because participants may not recall all public data, recall is a lower bound.

\item \textbf{Inferred intelligence.} Participants rate each inferred attribute and conversational topic (Sec.~\ref{subsec:profile-model}) on a 0--5 agreement scale, aggregated into per-category scores.

\item \textbf{Social-engineering engagement.} Drawing from the threat actor campaigns in Sec.~\ref{subsec:dual-use}, we define a 12-question ranking instrument that probes three operational problems every attack must solve, regardless of tactic: \textit{Career} (Q1--Q4), a credible pretext for first contact based on experience, skills, and career trajectory; \textit{Communication} (Q5--Q8), interaction calibration based on communication tone, OPSEC awareness, and language; and \textit{Interests} (Q9--Q12), rapport building based on personal interests, places visited, and recent activities. Each question presents four profile-derived options; participants rank them, and the system predicts the same ranking. Kendall's $\tau$~\cite{kendall1938}, compared against a random baseline (Wilcoxon signed-rank, $\alpha = 0.05$), measures how well the profile anticipates the target's choices, which is the operational core of a tailored attack (Sec.~\ref{subsubsec:results-quiz}). Questions and field mappings are provided in the Supplementary Material.

\item \textbf{Perception survey.} A structured questionnaire (nine 0--5 Likert items and three open-ended items) covers four areas and complements quantitative metrics; details are provided in the Supplementary Material. \textit{Context} (S1--S3) captures baseline social-media activity and privacy attitudes, contextualizing extraction against data availability. \textit{Accuracy} (S4--S6) measures self-recognition across extraction, inference, and engagement, indicating whether profiles felt authentic. \textit{Operational value} (S7--S9) captures perceived usefulness, including effectiveness against real targets, social-engineering potential, and professional adoption. \textit{Open feedback} (S10--S12) gathers strengths, weaknesses, and suggested improvements.
\end{itemize}

\section{Empirical Validation}
\label{sec:validation}

As described in Sec.~\ref{sec:methodology}, we validate the work on two complementary axes: the capability \emph{framework} and its \emph{operational} instantiation. The first establishes that the framework is coherent and grounded in real adversary behavior; the second, that the LLM-augmented pipeline turns it into accurate, exploitable profiles in practice.

\subsection{Framework Validation}\label{sec:framework_validation}

\definecolor{phbg}{RGB}{58,58,58}
\definecolor{heatred}{RGB}{220,140,140}
\definecolor{heatyellow}{RGB}{240,230,160}
\definecolor{heatgreen}{RGB}{160,210,160}
\definecolor{moved}{RGB}{200,218,248}

\begin{figure*}[!ht]
\centering
\caption{Expert panel validation ($n=13$). Relevance expressed as $\mu{\pm}\sigma$ [I-CVI]. Scale: 1 (not relevant) to 5 (highly relevant). I-CVI $\in [0,1]$, I-CVI $\geq 0.78$ acceptable. $\rightarrow$ phase reassigned.}
\label{fig:framework-validation}
\scriptsize
\setlength{\tabcolsep}{5pt}
\renewcommand{\arraystretch}{1.0}
\setlength{\arrayrulewidth}{2pt}
\arrayrulecolor{white}

\newcommand{\cvicolor}[1]{%
  \ifdim #1 pt < 0.5pt
    heatred!\fpeval{round((1-#1*2)*100,0)}!heatyellow%
  \else
    heatyellow!\fpeval{round((1-(#1-0.5)*2)*100,0)}!heatgreen%
  \fi
}

\newcommand{\V}[4]{%
  \ifx&#4&\cellcolor{\cvicolor{#3}}\else\cellcolor{moved}\fi%
  \parbox[c][0.75cm][c]{\linewidth}{%
    \centering\textbf{#1}\par
    \vspace{1pt}%
    {\scriptsize $#2$\enspace[#3]}%
  }%
}

\newcommand{\Vx}{\parbox[c][0.75cm][c]{\linewidth}{}}

\begin{tabular*}{\textwidth}{@{\extracolsep{\fill}}
  >{\centering}p{0.19\textwidth}|
  >{\centering}p{0.17\textwidth}|
  >{\centering}p{0.18\textwidth}|
  >{\centering}p{0.18\textwidth}|
  >{\centering\arraybackslash}p{0.17\textwidth}
@{}}
\rowcolor{phbg}
\textcolor{white}{\textbf{Planning \& Direction}} &
\textcolor{white}{\textbf{Collection}} &
\textcolor{white}{\textbf{Processing}} &
\textcolor{white}{\textbf{Analysis \& Corr.}} &
\textcolor{white}{\textbf{Dissemination}} \\\hline

\V{Strategy definition}{4.9{\pm}0.3}{1.00}{} &
\V{Passive observation}{4.9{\pm}0.3}{1.00}{} &
\V{Data triage \& filtering}{5.0{\pm}0.0}{1.00}{} &
\V{Behavior \& pattern}{4.9{\pm}0.3}{1.00}{} &
\V{Info sharing}{4.8{\pm}0.4}{1.00}{} \\\hline

\V{Risk assessment}{4.5{\pm}0.7}{0.92}{} &
\V{Automated acquisition}{4.9{\pm}0.3}{1.00}{} &
\V{Multi-channel normaliz.}{4.2{\pm}0.6}{0.92}{} &
\V{Media analysis}{4.1{\pm}0.5}{0.92}{} &
\V{Threat reporting}{4.9{\pm}0.3}{1.00}{} \\\hline

\V{Intel.\ requirements}{4.8{\pm}0.6}{0.92}{} &
\V{Active engagement}{4.0{\pm}0.7}{0.77}{} &
\V{KB ingestion}{4.2{\pm}0.8}{0.77}{} &
\V{People network}{4.9{\pm}0.3}{1.00}{} &
\V{Lessons learnt}{4.2{\pm}0.6}{0.92}{} \\\hline

\V{Legal \& ethical}{4.8{\pm}0.6}{0.92}{} &
\V{Multimedia collection}{4.2{\pm}0.7}{0.85}{} &
\V{Entity extraction}{4.7{\pm}0.5}{1.00}{} &
\V{Trend \& temporal}{4.4{\pm}0.7}{0.92}{} &
\V{Dashboard \& visual.}{3.8{\pm}1.1}{0.69}{} \\\hline

\V{Collection strategy}{4.7{\pm}0.6}{0.92}{} &
\V{Crowd-sourced collect.}{3.2{\pm}0.7}{0.23}{} &
\V{Identity resolution}{4.5{\pm}0.5}{1.00}{} &
\V{Intent \& deception}{4.5{\pm}0.5}{1.00}{} &
\V{Intel.\ liaison}{4.1{\pm}0.8}{0.77}{} \\\hline

\V{Digital identity}{4.2{\pm}0.6}{0.92}{} &
\V{Historical research}{4.0{\pm}0.6}{0.85}{} &
\V{Data enrichment}{4.3{\pm}0.5}{1.00}{} &
\V{Debriefing}{4.0{\pm}0.8}{0.69}{} &
\V{Oper.\ actionability}{4.8{\pm}0.4}{1.00}{} \\\hline

\V{Tool setup}{4.2{\pm}0.8}{0.77}{} &
\Vx &
\V{$\rightarrow$Metadata collect.}{4.1{\pm}0.5}{0.92}{m} &
\V{$\rightarrow$Attr.\ inference}{3.9{\pm}0.5}{0.85}{m} &
\Vx \\\hline

\V{Engagement planning}{4.4{\pm}0.5}{1.00}{} &
\Vx &
\Vx &
\V{$\rightarrow$Profile construction}{4.4{\pm}0.7}{0.92}{m} &
\Vx \\\hline

\Vx & \Vx & \Vx &
\V{Group dynamics}{4.2{\pm}0.6}{0.92}{} &
\Vx \\\hline

\Vx & \Vx & \Vx &
\V{Event \& threat det.}{4.8{\pm}0.4}{1.00}{} &
\Vx \\\hline

\Vx & \Vx & \Vx &
\V{Pivot discovery}{4.6{\pm}0.5}{1.00}{} &
\Vx \\

\end{tabular*}
\end{figure*}
\arrayrulecolor{black}
\setlength{\arrayrulewidth}{0.4pt}

An expert panel rated each capability for relevance, phase placement, and completeness to validate the framework. As shown in Fig.~\ref{fig:framework-validation}, experts confirmed the relevance of all 38 capabilities (overall $\mu = 4.42/5$, $\sigma = 0.40$). I-CVI, defined as the proportion of experts rating a capability $\geq 4$, exceeded the 0.78 content-validity threshold for 31 out of 38 capabilities, with an overall average of 0.90. No capability was flagged as superfluous by any expert.

Seven capabilities fell below the 0.78 I-CVI threshold, although six maintained high mean relevance ($\mu \geq 3.9$) and remained close to the acceptance boundary. Crowd-sourced Collection scored lower (I-CVI $= 0.23$, $\mu = 3.2$), reflecting limited operational familiarity among practitioners rather than conceptual irrelevance. All seven were retained for framework completeness.

Inter-rater agreement on phase classification, that is, the degree to which independent experts assign the same phase, was almost perfect (Fleiss' $\kappa = 0.92$), well above the 0.61 substantial-agreement threshold~\cite{landis1977}, with 95\% of assignments confirmed. \textit{Metadata Collection}, \textit{Attribute Inference}, and \textit{Profile Construction} received consistent reassignment suggestions from at least 5 of 13 experts and were moved accordingly.

Six experts suggested additional capabilities not covered in the original taxonomy, including operational exposure management, cross-discipline intelligence integration, and AI/LLM governance. These concepts were partially covered by existing categories, and the feedback was used to refine capability descriptions and align naming conventions.

The overall framework-structure assessment (1--5 Likert scale) confirmed that all 13 experts considered the framework to cover the full spectrum of social intelligence operations ($\mu = 4.2$), to provide sufficiently precise capability descriptions for operational use ($\mu = 3.8$), and to be suitable for adoption as a reference framework ($\mu = 4.5$).

\subsection{Operational Validation} \label{sec:results}
This section tests the operational value of the collected intelligence: we extract and infer participant profiles, assess quality against their own data as ground truth, and evaluate whether they can drive a credible social engineering attack.

\subsubsection{Pipeline Execution} \label{subsubsec:results-execution}

Across the 14 participants, execution time averaged 508s per participant ($\sigma = 31$s; $\approx$2h in total), indicating stable behavior regardless of profile complexity. Token consumption averaged 81.2K per participant ($\sigma = 16.6$K; 1.14M total), with higher relative variance (CV = 20.4\%) than execution time (CV = 6.1\%), reflecting heterogeneous profile content across participants. Cost impact remains negligible, with per-profile API cost between \$0.14 and \$0.18 (\$2.23 total). This is the full marginal cost per additional target, the threat-relevant metric for adversarial scalability, because the pipeline runs entirely on local infrastructure using open-source tooling (Langflow, Playwright), with no subscription or licensing costs beyond pay-per-use LLM API calls. The pipeline encountered 6 JSON parsing errors out of 126 API requests (4.8\%); all were automatically recovered via retry logic, and all runs completed successfully.

\subsubsection{Extracted Data} \label{subsubsec:results-extraction}

\begin{table}[pos=!b]
\centering
\caption{Extraction performance by field category. $n$ = participants with at least one item.}
{\footnotesize
\label{tab:extraction-category}
\begin{tabular}{l|c|rrr|rrr}
\toprule
Category & $n$ & TP & FP & FN & $P$\% & $R$\% & $F_1$\% \\
\midrule
Identity       & 14 &  44 & 4 &  0 & 91.7 & 100  & 95.7 \\
Experience     & 14 & 249 & 4 &  0 & 98.4 & 100  & 99.2 \\
Education      & 14 & 184 & 2 &  4 & 98.9 & 97.9 & 98.4 \\
Certifications & 12 & 120 & 0 & 16 & 100  & 88.2 & 93.8 \\
Projects       &  4 &  24 & 0 &  0 & 100  & 100  & 100  \\
Places Visited & 11 &  69 & 5 &  5 & 93.2 & 93.2 & 93.2 \\
Social Links   &  8 &  21 & 0 &  0 & 100  & 100  & 100  \\
\midrule
\textbf{Total} & & \textbf{711} & \textbf{15} & \textbf{25} & \textbf{97.9} & \textbf{96.6} & \textbf{97.3} \\
\bottomrule
\end{tabular}}
\end{table}

The pipeline achieved $F_1 = 97.3\%$ ($P = 97.9\%$, $R = 96.6\%$). Four participants scored a perfect $F_1$; the lowest score (84.1\%) came from the most data-rich profile (114k tokens), consistent with LLM degradation on long inputs~\cite{liu-etal-2024-lost}. This profile alone accounts for 20 of 25 false negatives. False positives remained low and distributed (max 5 per participant), so precision is a robust property of the pipeline, while recall is more sensitive to profile complexity.

By category (Tab.~\ref{tab:extraction-category}), most fields reach perfect or near-perfect $F_1$; only \textit{Certifications} ($F_1 = 93.8\%$) and \textit{Places Visited} ($F_1 = 93.2\%$) underperform: the former from context saturation in the richest profile (12 of 28 credentials extracted), the latter from locations embedded in photo metadata not reachable by the current scraper.

\subsubsection{Inferred Intelligence} \label{subsubsec:results-inferred}

\begin{table}
\centering
\caption{Inference agreement and confidence calibration ($n=14$). Rating: participant agreement (0--5). Conf: model confidence (0--5). Diff = Conf$-$Rating; negative = underconfidence.}
\label{tab:inference-category}
\resizebox{\columnwidth}{!}{%
\begin{tabular}{@{}ll rrrr rr@{}}
\toprule
 & & \multicolumn{4}{c}{Rating} & \multicolumn{2}{c}{Calibr.} \\
\cmidrule(lr){3-6} \cmidrule(lr){7-8}
Class & Category & $\mu$ & IQR & $\sigma$ & Min & Conf & Diff \\
\midrule
\multirow{5}{*}{\rotatebox{90}{Target}}
 & Short Bio              & 4.57 & 1.00 & 0.65 & 3   & 4.47 & $-$0.10 \\
 & Main Skills            & 4.79 & 0    & 0.58 & 3   & 4.40 & $-$0.39 \\
 & OpSec Awareness        & 4.79 & 0    & 0.58 & 3   & 3.94 & $-$0.84 \\
 & Tech.\ Sophisticat.  & 4.86 & 0    & 0.36 & 4   & 4.15 & $-$0.71 \\
 & Org.\ Access           & 4.57 & 0.75 & 0.76 & 3   & 4.19 & $-$0.38 \\
\midrule
\multirow{3}{*}{\rotatebox{90}{Interests}}
 & Prof.\ Interests       & 4.71 & 0    & 0.83 & 2   & 4.26 & $-$0.45 \\
 & Pers.\ Interests       & 4.64 & 0    & 0.74 & 3   & 3.98 & $-$0.66 \\
 & Recent Activities      & 4.79 & 0    & 0.43 & 4   & 3.92 & $-$0.86 \\
\midrule
\multirow{4}{*}{\rotatebox{90}{Behavior}}
 & Career Trajectory      & 4.86 & 0    & 0.36 & 4   & 3.31 & $-$1.54 \\
 & Geo.\ Mobility         & 4.71 & 0    & 0.61 & 3   & 3.78 & $-$0.94 \\
 & Comm.\ Tone            & 4.86 & 0    & 0.36 & 4   & 4.30 & $-$0.55 \\
 & Primary Lang.          & 4.79 & 0    & 0.58 & 3   & 4.24 & $-$0.55 \\
\midrule
\multirow{2}{*}{\rotatebox{90}{Eng.}} & \multirow{2}{*}{Conversational} & \multirow{2}{*}{4.55} & \multirow{2}{*}{0.33} & \multirow{2}{*}{0.78} & \multirow{2}{*}{2.8} & \multirow{2}{*}{3.73} & \multirow{2}{*}{$-$0.82} \\
 & & & & & & & \\
\midrule \midrule
\multicolumn{2}{@{}l}{\textbf{Overall}} & \textbf{4.73} & & \textbf{0.50} & & \textbf{4.05} & $\mathbf{-0.68}$ \\
\bottomrule
\end{tabular}}
\end{table}

Participants rated 18 inferred attributes on a 0--5 Likert scale: 12 attributes plus six conversational topics per participant ($18 \times 14 = 252$ total ratings). The overall mean is 4.73 ($\sigma = 0.50$), with 79.1\% of ratings at the maximum and a median of 5 in every category. Ten categories show near-unanimous agreement (IQR = 0), while the other three (\textit{Short Bio}, \textit{Organizational Access}, \textit{Conversational Topics}) show moderate spread. Two participants account for 11 of the 13 ratings below 4, pointing to self-perception gaps rather than systematic inference failures. Reported scores therefore represent a conservative estimate of actual inference accuracy. Per-category agreement and calibration are shown in Tab.~\ref{tab:inference-category}.

Across all 13 categories, model self-confidence (Conf) falls below participant ratings: the Diff column (Conf\ $-$ Rating) is uniformly negative (overall mean $-0.68$), indicating systematic underconfidence. The gap is smallest for factual attributes read directly from profile fields (\textit{Target} class, mean Diff $-0.48$) and largest for attributes requiring cross-entry reasoning (\textit{Behavior} class, mean Diff $-0.90$, driven by \textit{Career Trajectory} at $-1.54$). These scores should therefore be interpreted as relative reliability indicators for operational triage rather than calibrated probabilities.

\subsubsection{Social Engineering Engagement Assessment} 
\label{subsubsec:results-quiz}

The engagement assessment evaluates whether extracted and inferred attributes are sufficient to personalize a social engineering interaction, measured by Kendall's $\tau$ ($-1$: inverse match, $0$: no correlation, $+1$: exact match).

\begin{figure}[pos=b]
\centering
\includegraphics[width=0.95\linewidth]{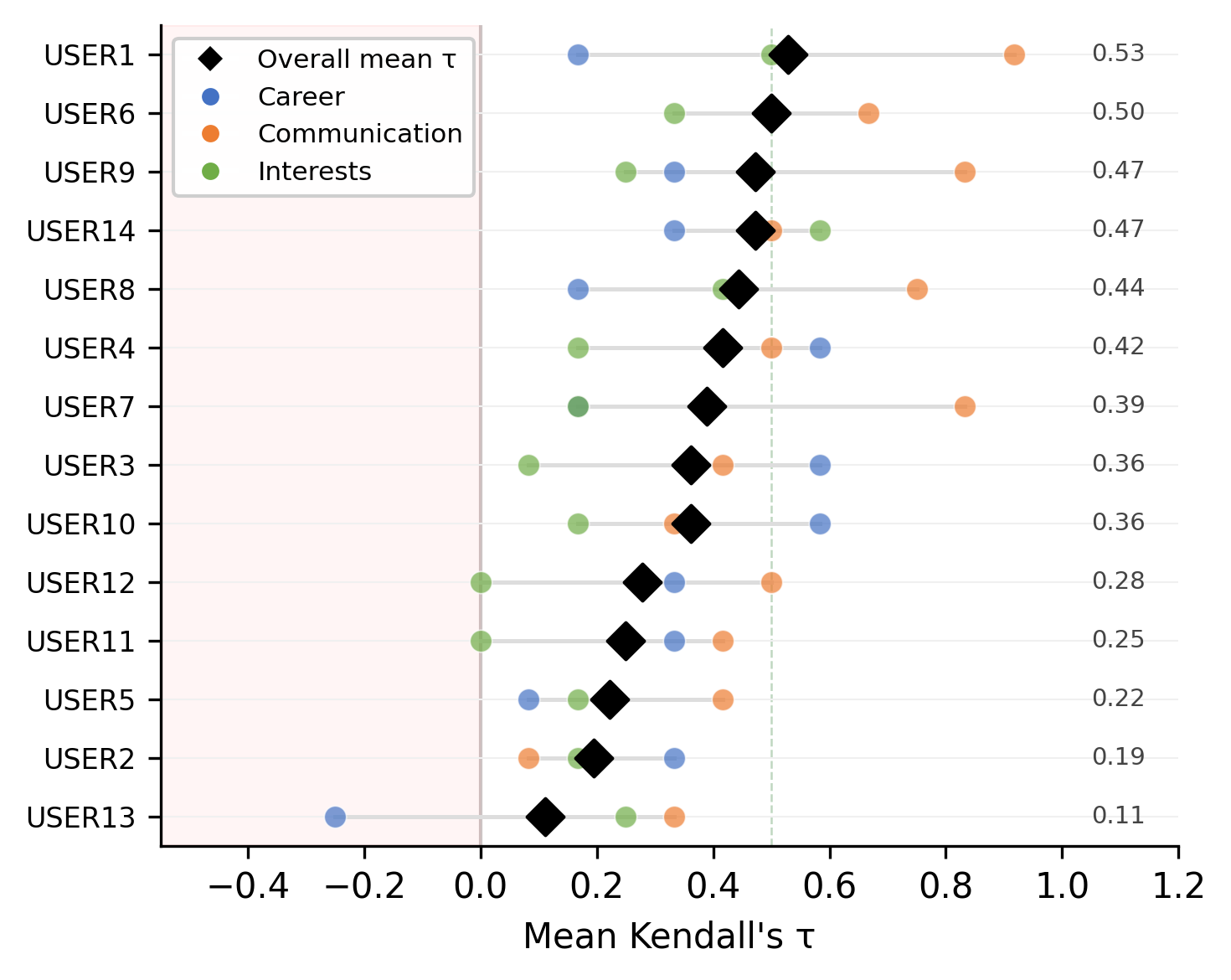}
\caption{Engagement assessment: per-subject profiling accuracy. Diamonds: mean $\tau$; dots: per-category means.}
\label{fig:quiz-dotplot}
\end{figure}

The overall mean $\bar{\tau} = 0.36$ ($\sigma = 0.51$, median $= 0.33$) indicates moderate agreement between predicted and observed rankings, with 71\% of observations showing positive $\tau$ and 15\% achieving exact matches. The proportion of questions where the system correctly predicts the participant's first choice (Top-1 accuracy) is about 52\%, doubling the random baseline. All 14 per-subject means are strictly positive (including participants with limited social-media footprints), suggesting that even partial digital traces provide sufficient signal for non-trivial inference (Fig. \ref{fig:quiz-dotplot}). A Wilcoxon signed-rank test on per-subject $\bar{\tau}$ values confirms that overall agreement is significantly above chance ($W = 105$, $p < 0.001$, $n = 14$).

\begin{figure}
\centering
\includegraphics[width=1.0\linewidth]{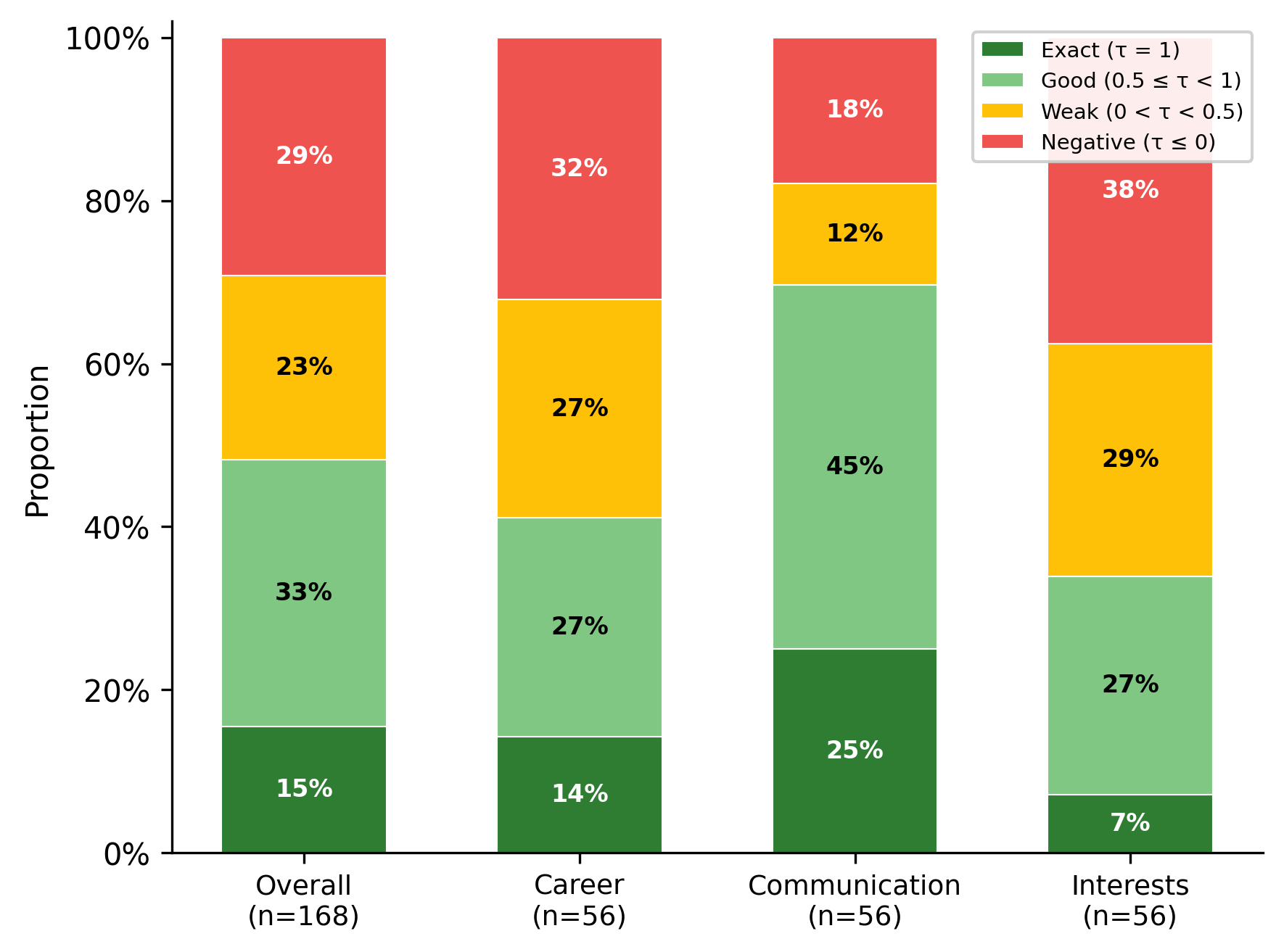}
\caption{Engagement assessment: ranking quality distribution by question category.}   
\label{fig:quiz-stacked}
\end{figure}

Communication achieves the highest $\bar{\tau}$ (0.54), followed by Career (0.30) and Interests (0.23). In quality-tier terms (Fig.~\ref{fig:quiz-stacked}), Communication yields 70\% Good or Exact predictions ($\tau \geq 0.5$), compared with 41\% for Career and 34\% for Interests. This asymmetry reflects the nature of the underlying data: communication-related fields are derived from recurrent behavioral patterns (writing style, language choice, disclosure habits) that tend to remain stable across posts. Interest-related fields require aggregation of sparse, context-dependent signals. Career falls between the two: professional attributes are often explicitly stated, but their preference implications still require additional inference. This hierarchy suggests that interaction calibration is the most accurately personalizable attack dimension from retrieved data in this sample, while rapport-based approaches require richer behavioral signals than two-platform collection currently provides.

\subsubsection{Perception Survey}
\label{subsubsec:results-survey}

\begin{figure}[pos=t]
\centering
\includegraphics[width=\columnwidth]{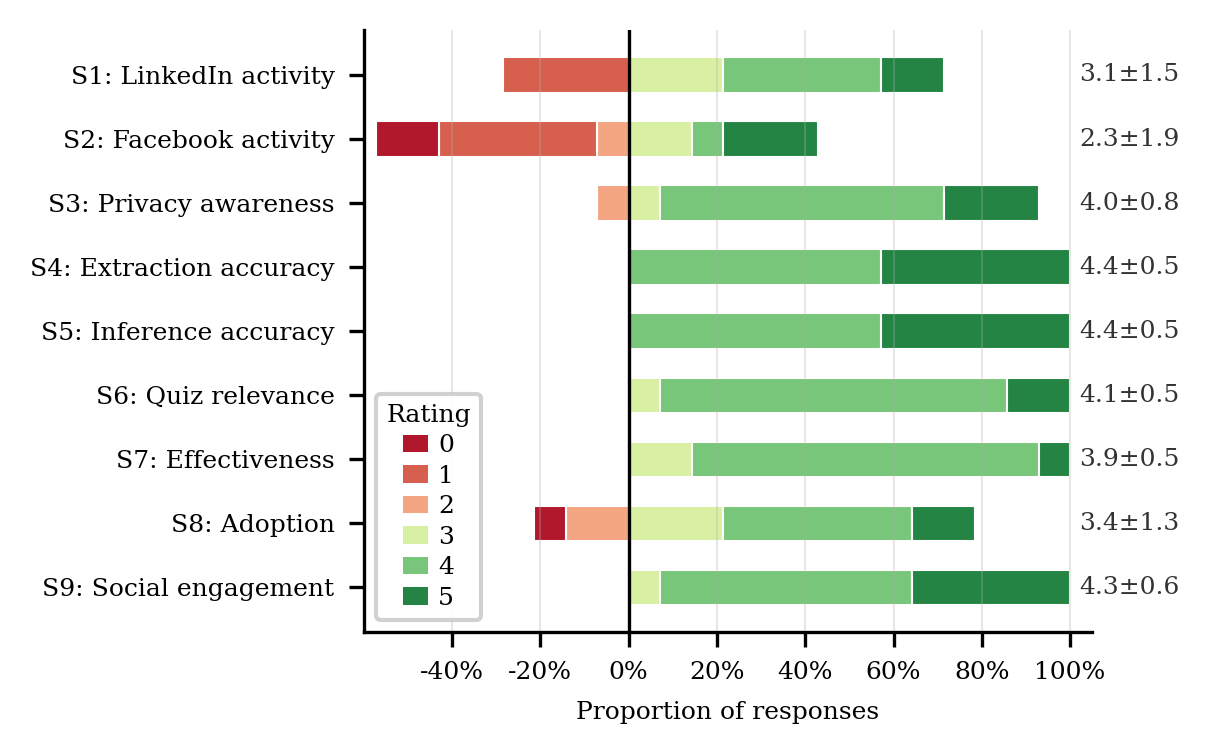}
\caption{Perception survey: likert-scale responses (0--5, $n = 14$). Bars extend left for ratings $\leq 2$ and right for ratings $\geq 3$.}
\label{fig:survey-likert}
\end{figure}

As Fig.~\ref{fig:survey-likert} shows, pipeline outputs were perceived as highly accurate even by privacy-conscious users with varying activity levels. Indeed, despite variance in social-media activity across participants (S1--S2) and uniformly high privacy awareness (S3), participants rated extraction and inference accuracy highly (S4--S5, $\mu = 4.4$, with all 14 ratings at 4 or 5) and found the engagement questions relevant to their real profiles (S6, $\mu = 4.1$).

Participants themselves acknowledge that the generated profile could support a convincing social engineering attempt: perceived effectiveness against a real target (S7, $\mu = 3.9$) and social-engagement potential (S9, $\mu = 4.3$, 93\% of ratings at 4 or 5) both score high. Professional adoption (S8), by contrast, shows substantial disagreement ($\mu = 3.4$, $\sigma = 1.3$), likely reflecting differences in professional role rather than doubts about effectiveness.

Open-ended responses reinforce the picture. The most-cited strength (S11) was overall profile accuracy, with two participants reporting heightened awareness of their own privacy exposure on social media; the main weaknesses (S10) were overfitting, outdated content, and incorrect collected data (each reported by 29\% of participants), followed by missing data and limited platform coverage. This aligns with suggested improvements (S12): additional data sources (64\%), broader platform integration (21\%), and deeper reasoning capabilities (36\%, including improved explainability, temporal inference, and a broader engagement instrument).

\section{Discussion}
\label{sec:discussion}

The framework's capabilities are inherently dual-use, differing only in intent and authorization. Attackers wield them against organizations (data breaches, insider recruitment, service disruption, financial fraud), the public (electoral interference, reputation attacks), and individuals (doxing). Defenders run the same capabilities the other way, assessing their own exposure, monitoring open and crowd-sourced signals for early warning, countering influence operations, and profiling the adversaries themselves.

At $\sim$\$0.16 per target with no human intervention (under \$20 for a hundred), the cost and expertise barriers to large-scale profiling have collapsed~\cite{hazell2023}. The economics cut both ways: defenders can self-assess their exposure before adversaries strike, while low-resource attackers gain what once required well-funded operations. As profiling turns cheap and automatic, exposure becomes an operational certainty, making a shared framework to assess and bound it urgent.

No single source we surveyed grounds more than roughly two-thirds of the 38 capabilities (at most $\sim$68\%), and only their union spans the full set. Our taxonomy is the first to provide this integration: prior systems cover only one stage, such as threat-report extraction, isolated attribute inference, or attack-content generation. Our pipeline instead delivers a complete, traceable, and ground-truth-validated profile.

\subsection{Baseline comparison} \label{subsubsec:baselinecomparison}

To quantify the LLM's contribution over conventional methods, we ran two baselines on the same raw scrapes, scored against the 736 validated ground-truth items. The regex baseline targets syntactically regular fields (temporal expressions, emails, URLs); the spaCy baseline uses the \textit{en\_core\_web\_lg} NER model, collecting entities labeled \textit{PERSON}, \textit{ORG}, \textit{GPE}, \textit{LOC}, \textit{DATE}, and \textit{FAC}. Matching is deliberately permissive (case-insensitive substring with stopword and short-token filtering), so baseline scores are an upper bound.

Both baselines fall far short (Tab.~\ref{tab:regex_spacy_compare}): regex reaches $F_1 = 28.0\%$, and spaCy NER reaches $17.3\%$. Although spaCy recovers more items, it also produces many false positives. Our pipeline reaches $97.3\%$. Neither baseline produces any of the 252 inferred attributes, which require reasoning across multiple profile fields.

Three structural differences explain this gap. First, the LLM performs semantic parsing: it decomposes a single text field into multiple typed attributes mapped to schema positions, whereas regex and spaCy detect entities in isolation and ignore their relational role in the profile. Second, its output is operationally precise and directly usable by an analyst, while baseline output requires extensive manual filtering that negates automation benefits. Third, only the LLM infers latent attributes across fields, turning the pipeline from an extraction tool into a profiling system. 

Within this evaluation, neither baseline approaches the pipeline's ability to extract structured records and generate actionable intelligence, supporting the LLM as the core processing component.

\subsection{Limitations and Future Directions} \label{subsec:future} 

\begin{table}[pos=b]
\centering
\footnotesize
\caption{Baseline comparison. Ground-truth: $n=736$ items (TP+FN). Inferred items (252) excluded as unreachable by both baselines.}
\label{tab:regex_spacy_compare}
\begin{tabularx}{\columnwidth}{@{}>{\raggedleft\arraybackslash}X| rrr| rrr@{}}
\toprule
\textbf{Method} & \textbf{TP} & \textbf{FP} & \textbf{FN} & \textbf{$P$\%} & \textbf{$R$\%} & \textbf{$F_1$\%} \\
\midrule
Regex           & 171 & 315 & 565 & 35.2 & 23.2 & 28.0 \\
spaCy NER       & 536 & 4{,}930 & 200 & 9.8 & 72.8 & 17.3 \\
\rowcolor{gray!20} LLM pipeline & 711 & 15 & 25 & 97.9 & 96.6 & 97.3 \\
\bottomrule
\end{tabularx}
\end{table}

\emph{Scope and generalizability.} By design, this is a feasibility pilot involving 14 cybersecurity professionals recruited within a professional network: the sample establishes end-to-end viability and yields performance significantly above chance ($p < 0.001$). A larger, more diverse cohort would generalize the results, at low marginal cost per target. Coverage is likewise bounded to Facebook/LinkedIn platforms and a single snapshot: broadening to further platforms and OSINT sources (e.g., Instagram, GitHub, breach databases) and to longitudinal capture would extend completeness and enable deeper behavioral tracking.

\emph{Validation and model dependence.} The high agreement on inferred attributes ($\mu = 4.73$) rests on participant self-report. This is the natural ground truth for one's own profile, although independent validation would reduce self-assessment bias. Residual hallucination remains (15 extraction false positives; 29\% of participants reported incorrect data), but it stays at operationally manageable levels ($P = 97.9\%$). Results also depend on specific model versions: benchmarking across LLMs would separate framework-level effects from model-specific effects, and chunking or summarization could mitigate accuracy loss in the densest profiles near context limits.

\emph{Operational extension.} Active engagement was assessed by proxy and not executed directly. A possible next step is controlled engagement under ethics-board oversight, ranging from LLM-driven conversational agents to assisted simulations, with an ablation that compares LLM-personalized and conventional non-tailored social engineering. The pipe\-line's low marginal cost also enables stress-testing platform defenses and developing reactive and proactive countermeasures against LLM-augmented reconnaissance, including active deception (honeytoken accounts and synthetic footprints). Integration with operational CTI platforms (MISP, OpenCTI, STIX/TAXII) and multimodal inputs (including metadata) would bridge the gap from prototype to deployable asset. Finally, the architecture exercises only a subset of the 38 capabilities: Sec.~\ref{sec:llm-augmentation} maps additional augmentation opportunities across all five CTI phases that remain to be explored.

\section{Conclusions} \label{sec:conclusions}

As adversaries automate target profiling at scale and negligible cost, manual defense is structurally outpaced. Anticipating individuals' exposure before an adversary exploits it now requires defenders to adopt the same automation.

To make that measurable, we unified SOCMINT and Cyber-HUMINT into a single expert-validated taxonomy and operationalized its core as an LLM-augmented pipeline under human oversight. On real users it showed that the exposure that matters is not only what a person posts but also what a machine can infer from it: beyond near-perfect extraction ($F_1 = 97.3\%$, far above rule- and NER-based methods), it derived operational, attack-relevant attributes the targets never disclosed, such as OPSEC awareness and organizational access, yet confirmed as accurate ($4.73/5$).

These profiles proved not merely descriptive but attack-ready: they anticipated targets' engagement preferences above chance, and the targets themselves judged them sufficient for a credible social-engineering approach. If a low-cost automated system reaches this level, an adversary can as well; organizations therefore cannot leave their own exposure unmeasured.

Our framework turns dual-use mirroring into a systematic, reusable self-assessment capability that an organization can run proactively to gauge its users' exposure and take countermeasures before an adversary exploits it. By mapping the augmentation surface across all five CTI phases, we expect this work to lay a foundation for the systematic study of LLM-augmented intelligence across the full observation-to-engagement continuum.

\section*{Ethical Considerations} 
\label{sec:ethics}

The framework treats SOCMINT and Cyber-HUMINT as legitimate dual-use CTI activities. This study is strictly defensive, aiming to advance responsible understanding of social-media exposure and inform countermeasures. It was unanimously approved by the Ethics Committee for Research with Human Subjects in Non-Biomedical Fields (CERSUB) of the University of Naples Federico II on 24 February 2026. The pipeline processes only content that participants made publicly visible, through authenticated sessions, with no deception, manipulation, or covert observation. All participants gave written informed consent covering objectives, procedures, instruments, retention, pseudonymization, and GDPR rights. Participation was voluntary, uncompensated, independent of any professional relationship, and withdrawable without consequence.

Data is pseudonymized during processing and anony\-mized for publication; raw captures and generated profiles are stored on encrypted, access-controlled devices, and all processing, including transmission to LLM endpoints, is GDPR-compliant. Raw data is retained for twelve months for reproducibility. The profile review also served as a privacy-awareness exercise, a benefit that the ethics committee recognized in its risk-benefit assessment.

\section*{Funding}
This research did not receive any specific grant from funding agencies in the public, commercial, or not-for-profit sectors.

\section*{Data availability}
The data collected from human participants contain sensitive and personally identifiable information. Due to privacy and ethical constraints, raw data are not publicly available; aggregated, de-identified data can be provided by the corresponding author upon reasonable request. The same applies to the Langflow pipeline configuration and scripts, which can be shared for research purposes. Full replication requires authenticated access to social media platforms and LLM API credentials, which cannot be redistributed. Prompt details are provided as Supplementary Material to support reproducibility.

\section*{Declaration of generative AI and AI-assisted technologies in the manuscript preparation process}
During the preparation of this work, the authors used Claude (Anthropic) for language editing, structural revision, clarity improvements, and coding support. Additionally, OpenAI models were used as part of the experimental methodology described in the paper. After using these tools, the authors reviewed and edited all content and code, and take full responsibility for the published article. 

\section*{Acknowledgments}
We thank the cybersecurity professionals who participated in the pilot study for their time and feedback. We are also grateful to Dario Colantoni, Stefano Battista, and Gennaro Caccavale for their valuable support during the development of this work.

\printcredits

\bibliographystyle{unsrtnat}
\bibliography{BIBLIO_social_humint_v4_0}

\end{document}